\documentstyle[12pt,epsf]{article}
\newcommand{\nc}{\newcommand}

\nc{\nn}{\nonumber}
\nc{\beq}{\begin{equation}}
\nc{\eeq}{\end{equation}}
\nc{\beqa}{\begin{eqnarray}}
\nc{\eeqa}{\end{eqnarray}}
\nc{\pz}{\partial_z}
\def\Slash#1{#1\kern-0.55em\raise.05ex\hbox{/}}
\def\slash#1{#1\kern-0.5em\raise.05ex\hbox{{$\scriptstyle /$}}}

\nc{\lsim}{\mbox{\raisebox{-.6ex}{~$\stackrel{<}{\sim}$~}}}
\nc{\gsim}{\mbox{\raisebox{-.6ex}{~$\stackrel{>}{\sim}$~}}}

\nc{\sss}{\scriptscriptstyle}
\nc{\bi}{\bibitem}

\def\ra{{\rightarrow}}

\def\rd{{\rm d}}

\def\bvzi{{\bar v_{i,z}}}

\def\ssk{{\sss (k)}}
\def\sumxi{{\sum_j\xi^\ssk_j}}

\def\Re{{\rm Re\,}}
\def\Im{{\rm Im\,}}
\def\diag{{\rm diag}}

\def\ie{{\em i.e.\ }}

\def\mhalf{{\scriptstyle \frac{1}{2}}}

\def\nBR#1{{\Bigl( #1 \Bigr)}}
\def\BBR#1{{\Bigl[ #1 \Bigr]}}

\def\NBR#1{{\left( #1 \right)}}
\def\ave#1{{\left\langle #1 \right\rangle}}
\def\aves#1{{\langle #1 \rangle}}

\def\scp{s_{\sss CP}}
\def\acp{\alpha_{\sss CP}}
\def\Aso{{A_1}}
\def\Ast{{A_2}}
\def\Uso{{U_1}}
\def\Ust{{U_2}}
\def\sfrac#1#2{{\textstyle{#1\over #2}}}
\begin{document}
%
%
%
\begin{titlepage}
\pagestyle{empty}
\baselineskip=21pt
\rightline{McGill 00-15}
\rightline{NORDITA 2000/38 HE}
\rightline{LPT-ORSAY 00-46}
\rightline{hep-ph/0006119}
\vskip .4in
\begin{center}
        {\Large{\bf Supersymmetric Electroweak Baryogenesis}}
\end{center}
\vskip .1in
\begin{center}
        James M.~Cline\\
          {\it McGill University, Montr\'eal, Qu\'ebec H3A 2T8, Canada}\\
          Michael Joyce \\
          {\it LPT, Universit\'e Paris-XI, B\^atiment 211,
               F-91405 Orsay Cedex, France}\\
Kimmo Kainulainen\\
          {\it  NORDITA, Blegdamsvej 17, DK-2100, Copenhagen \O ,
          Denmark\\
}
\vskip .2in
\end{center}
\baselineskip=18pt

We re-examine the generation of the baryon asymmetry in the minimal
supersymmetric standard model (MSSM) during the electroweak phase
transition. We find that the dominant source for baryogenesis
arises from the chargino sector. The CP-violation comes from the
complex phase in the $\mu$ parameter, which provides CP-odd contributions
to the particle dispersion relations. This leads to different accelerations
for particles and antiparticles in the wall region which, combined with
diffusion, leads to the separation of
Higgsinos and their antiparticles in the front of the wall. These asymmetries
get transported to produce perturbations in the left-handed chiral quarks,
which then drive sphaleron interactions to create the baryon asymmetry. We
present a complete derivation of the semiclassical WKB formalism, including
the chargino dispersion relations and a self-consistent derivation of the
diffusion equations starting from semiclassical Boltzmann equations for
WKB-excitations. We stress the advantages of treating the transport
equations in terms of the manifestly gauge invariant physical energy and
kinetic momentum, rather than in the gauge variant canonical variables
used in previous treatments. We show that a large enough baryon asymmetry
can be created for the phase of the complex $\mu$-parameter as small as
$\sim 10^{-3}$, which is consistent with bounds from the neutron
electric dipole moment.
\end{titlepage}
\newpage
%
%
%
%
\baselineskip=20pt

\section{Introduction}

It is a fascinating possibility that the baryon asymmetry of the universe
(BAU) may have been generated at the electroweak epoch (for reviews, see
\cite{REVS}).  The great attraction of this idea is that, in contrast to
other mechanisms operating at higher energy scales, it involves physics
which is being searched for at accelerators now.  An {\it a priori}
calculation of the baryon asymmetry, as accurate as that of the abundance
of the light elements in nucleosynthesis, may still be unattainable at
present, but we should nevertheless strive to compute it as carefully as
possible.  One hopes thereby to reach a definitive conclusion as to the
feasibility, at least, of generating the BAU at the electroweak scale.

While there are many theoretical motivations for considering extensions of
the standard model (SM), in the present context we are also prompted to
do so for the simple reason that the SM by itself appears unable to produce
the observed BAU. The smallness of the CP violation in the KM matrix appears
to be in itself an insurmountable obstacle to baryogenesis in the SM
(although there has been considerable debate on this subject \cite{CPDEB}),
and has motivated many studies of baryogenesis in extended models with
additional CP violation leading to more efficient baryon production.
In addition to this problem, moreover, the SM fails badly with respect to the
sphaleron wash-out bound\footnote{As discussed in \cite{kination}
this bound is predicated on the assumption that the Universe is
radiation dominated at the electroweak epoch, and can be significantly
weakened in non-standard (e.g. scalar field dominated) cosmologies.}.
Lattice studies have shown that for any value of the higgs mass, even well
below the present experimental lower limit, the phase transition would be
so weak that sphaleron interactions remain in equilibrium in the broken
phase of the electroweak sector, causing the baryon asymmetry to relax back
to essentially zero immediately after its generation \cite{KLRS1}.

Several extensions of the SM have been considered to overcome the sphaleron
wash-out bound by strengthening the phase transition \cite{AHTZ,JPT,JPT2,CKV}.
Best motivated from the particle physics point of view is the minimal
supersymmetric standard model (MSSM). Several recent perturbative and
nonperturbative studies of the properties of the phase transition in this
model \cite{old,new,LR,CM} have shown that in a restricted part of
the parameter space, the sphaleron bound can be satisfied. An important
question is therefore whether for these same parameter values the
{\it generation} of the observed BAU is possible.

Baryogenesis in the MSSM has already been studied in several papers
\cite{CN,HN,CKN,AOS,CQRVW,CJK,RS,CK,KK,HS}. The overall framework of the
baryogenesis mechanism is essentially agreed upon: bubbles nucleate
at a first order phase transition and the expanding bubble walls propagate
through the hot plasma, perturbing the quasiparticle distributions from
equilibrium in a CP-violating manner.  Incorporating the effects of
transport leads to a local excess or deficit of left-handed fermions over
their antiparticles on and around the propagating bubble walls. This drives
the anomalous baryon number violating processes to produce a net baryon
asymmetry, which is swept behind the bubble wall where it is frozen in
(assuming that the sphaleron bound is satisfied). Moveover,
common to all methods is reducing the problem to a set of diffusion
equations coupling the sourced species to the species that bias the
sphalerons. These are coupled equations which have the general form
\beq
	D_i \xi_i'' + v_w \xi' + \Gamma_i(\xi_i + \xi_j + \cdots) = S_i\;,
\eeq
where $i$ labels the particle species and $\xi=\mu_i/T$ is its chemical
potential divided by temperature. Primes denote spatial derivatives in
the direction ($z$) perpendicular to the wall, $v_w$ is the wall velocity,
$\Gamma_i$ is the rate of an interaction that converts species $i$ into
other kinds of particles, and $S_i$ is the source term associated with the
current generated at the bubble wall. There is little controversy about the
form of these equations, but little agreement exists as to how to properly
derive the source terms $S_i$. There are many different formalisms for
obtaining the sources \cite{HuSa,JPT,Riotto}, but so far little effort
has been made to see how far they agree or disagree with each other. We
shall comment on this issue briefly in our conclusions.

Here we shall use the `classical force' mechanism (CFM) for baryogenesis
\cite{JPT}, \cite{CJK,CK,KK}. The CFM makes use of the intuitively simple
picture of particles being transported in the plasma under the influence
of the classical force exerted on them by the spatially varying Higgs
field condensate. We assume that the
plasma in this bubble wall region can be described by a collection of
semiclassical quasiparticle states which we shall refer to as WKB states,
because their equation of motion is derived using the WKB approximation
expanding in derivatives of the background field.  The force
acting on the particles can be deduced from the WKB dispersion relations and
their corresponding canonical equations of motion. This is a reasonable
assumption when the de Broglie wavelength of the states is much shorter than
the scale of variation of the bubble wall, \ie $\lambda \ll \ell_w$, which is
satisfied in electroweak baryogenesis; in the MSSM, the wall widths are
typically $\ell_w \sim 6-14/T$ \cite{CM,MQS}, whereas for a typical
excitation $\lambda \sim 1/T$. Given these conditions one can write a
semiclassical Boltzmann equation for the distribution functions of the
local WKB-states
\beq
        (\partial_t + {\bf v}_g \cdot\partial_{\bf x}  +
        {\bf F} \cdot\partial_{\bf p}) f_i = C[f_i,f_j,...].
\label{SCBE}
\eeq
where the group velocity and classical force are given respectively by
\beq
\label{GVF}
	{\bf v}_g \equiv \partial_{{\bf p}_c}\omega;\qquad
	{\bf F} = \dot {\bf p} = \omega \dot {\bf v}_g.
\eeq
Here ${\bf p}_c$ is the canonical, and ${\bf p} \equiv \omega {\bf v}_g$ the
physical, kinetic momentum along the WKB worldline. Note that we treat the
transport problem here in the  kinetic variables - in which the
Boltzmann equation has the non-canonical form of (\ref{SCBE}) -  rather
than in the canonical variables used in previous treatments. As will be
discussed in more detail below, this choice has the simple advantage of
circumventing all the difficulties associated with the variance of the
canonical variables under local phase (`gauge') transformations of the fields
in the Lagrangian. In these kinetic variables it is also more manifestly
(and gauge independently) clear how, because of CP-violating effects,
particles and antipartices experience different forces in the wall
region,
which leads to the separation
of chiral currents. The explicit form of ${\bf v}_g$ and ${\bf F}$ in a given
model can be found from the WKB dispersion relations, as we will illustrate
in sections 2 and 3. The Boltzmann equation (\ref{SCBE}) can then be converted
to diffusion equations in a standard way by doing a truncated moment expansion
\cite{CJK} (see section 4).

\vskip 0.3cm
The largest contribution to baryogenesis in the MSSM comes from the chargino
and neutralino sectors.  For the charginos, the
CP violating effects are due to the complex parameters $m_2$ and $\mu$ in the
mass term
\beq
\label{chmass}
	\bar\psi_R M \psi_L = (\overline{\widetilde w^{^+}},\
	\overline{\widetilde h^{^+}}_{1} )_{R}
	 \left(\begin{array}{cc} m_2 & g H_2 \\
	                          g H_1 & \mu
	         \end{array}\right)
	 \left(\begin{array}{c}
	             \widetilde w^{^+} \\
	             \widetilde h^{^+}_{2}
	         \end{array}\right)_{\!\!L}.
\eeq
The complex phases, combined with the mixing due to the Higgs fields, which
vary inside the bubble wall, give rise to spatially varying effective phases
for the mass eigenstates, which induce CP-violating currents for these
excitations.  To get analytic results, one can try to compute the current to
leading order in an expansion in derivatives of the Higgs fields. This is the
procedure followed in all methods designed to work on the thick wall limit
\cite{HuSa,JPT,Riotto,CJK}. This approximation cannot be used in the quantum
reflection case \cite{JPT2,AOS,CKV,RS}, which can be relevant in the limit
of very thin bubble walls.

We comment here on an apparent discrepancy in the literature concerning
the derivative expansion of the chargino source.  References \cite{HN} and
\cite{CQRVW} obtained a source for the $H_1-H_2$ combination of Higgs
currents of the form
\beq
\label{msign}
         S_{H_1-H_2} \sim {\rm Im}(m_2\,\mu)\,(H_1 H_2' - H_2 H_1'),
\eeq
whereas ref.\ \cite{CJK} found the other orthogonal linear combination,
$H_1 H_2' + H_2 H_1'$.  We previously believed that the disagreement
was because of fundamental differences between our CFM formalism and those
of refs. \cite{HN,HuSa,CQRVW,Riotto}.  However we recently understood
\cite{CK,KK} that the difference was partially due to the fact that we
were in fact computing the source for $H_1 + H_2$, for which the
result is
\beq
\label{psign}
         S_{H_1+H_2} \sim {\rm Im}(m_2\,\mu)\,(H_1 H_2' + H_2 H_1'),
\eeq
Therefore the disagreement about the sign was spurious: it can be shown
that all three methods actually agree with eq.\ (\ref{psign}); it simply
was not computed by the other references \cite{HN,CQRVW,Riotto,RS}.

The reason that the combination $H_1+H_2$ was not considered by other
authors is that it tends to be suppressed by Yukawa  and
helicity-flipping interactions from the $\mu$ term in the chargino mass
matrix.  Let us define chemical potentials for $H_1$, $H_2$, left-handed
third generation quarks $q_3$ and right-handed top quarks $t$, which we
will assume are equal to the chemical potentials for the corresponding
supersymmetric partners, as a consequence of supergauge interactions
mediated by gauginos.
If all the interactions arising from the Lagrangian
\beqa
          V &=&   \mu\tilde h_1 \tilde h_2
              + y h_2 \bar u_R q_L
              + y \bar u_R \tilde h_{2L}\tilde q_L
              + y \tilde u^*_R \tilde h_{2L} q_L
\nonumber\\
            &-& y \mu h_1 \tilde q^*_L \tilde u_R
              + yA_t\tilde q_L h_2\tilde u^*_R
              + \hbox{h.c.},
\label{Vint}
\eeqa
were considered to be in thermal equilibrium, they would give rise to
the constraints $\mu_{H_1} - \mu_{Q_3} + \mu_T = 0$, $\mu_{H_2} +
\mu_{Q_3} - \mu_T = 0$ and $\mu_{H_1} + \mu_{H_2} = 0$.  If these
conditions hold, the effect of the source $S_{H_1+H_2}$ is clearly
damped to zero. However, the rates of the processes coming from
(\ref{Vint}) are finite, and by studying the diffusion equations one
can show that there are corrections of order $(D_h\Gamma)^{-1/2}\sim
1$, where we used $D_h \simeq 20/T$ and the Yukawa rate $\Gamma \simeq
0.02 T$ (see eq.\ (\ref{stdset}) and the discussion following).

Even in formalisms where the source $S_{H_1-H_2}$ is nonvanishing
\cite{CQRVW,Riotto,RS}, one should then not neglect the source
$S_{H_1+H_2}$ without first checking whether the other source, $S_{H_1-H_2}$
really gives a larger effect. In fact $S_{H_1-H_2}$ does suffer from a severe
suppression:  quantitative studies of the electroweak bubbles in the MSSM
show that the ratio $H_2/H_1$ remains nearly constant inside the bubble walls
\cite{MQS,CM}; in Monte Carlo searches of the MSSM parameter space, the
deviation from constancy is typically at the level of one part in $10^3$, and
never more than $0.02$. Therefore the source $S_{H_1-H_2}$ is suppressed from
the outset by a factor of $10^2-10^3$ relative to $S_{H_1+H_2}$, which is
much worse than the Yukawa equilibrium suppression estimated above.

In the CFM the situation concerning the source $S_{H_1-H_2}$ is even
worse: we will show that there will be no source arising from classical
force, when computed correctly.
To see this is actually quite subtle, and relates to the question of the
gauge invariance we have referred to. If the problem is considered solely
in terms of the canonical variables, there appears to be a non-trivial
source of the form (\ref{msign}). That this term is unphysical however,
is indicated by the fact that it can be transformed away by a field
redefinition of the form
\beq
         \widetilde h^{^+}_{iL} \to e^{i\alpha_i} \widetilde h^{^+}_{iL},
\eeq
where $\alpha_i \sim \int {\rm Im}(m_2\mu)(H_1'H_2-H_2'H_1) dz$. Below we
will see that no such field redefinition has any effect on the physical
momenta or currents, and hence should not give rise to a physical force
(see also \cite{KK}). In our treatment in terms of the gauge invariant
kinetic variables this result is evident. In particular then the new
source for baryogenesis in the CFM picture found in \cite{HS} is absent
in our treatment.

Our main result is that baryogenesis remains viable for a large part of the
MSSM parameter space, possibly with the explicit CP-violating phase as small
as ${\rm arg}(m_2\mu) \sim 10^{-3}$. The efficiency depends on the assumed
squark spectrum, and the strongest baryoproduction corresponds to the light
right-handed stop scenario, which is also independently favored by the
sphaleron wash-out constraint \cite{new,LR}. The resulting asymmetry has a
complicated dependence on the wall velocity and for some parameters it
peaks around the value of $v_w \simeq 0.01$ which has been indicated by
recent studies of $v_w$ \cite{Moore,JS}.

\vskip 0.3cm The rest of the paper is structured as follows. In section 2 we
consider the simple case of a Dirac fermion with a complex spatially varying
mass. We determine the dispersion relation for the two helicities to leading
order in Higgs field derivatives, and find from it the group velocity and the
physical force acting on a fermion.  We also compute and interpret the
currents in the absence of collisions and show explicitly how the gauge
invariant force can be identified from canonical equations of motion.  In
section 3 we employ the formalism in the case of the MSSM, in particular,
computing the dispersion relations, group velocities and force terms for the
charginos. (Squarks and neutralinos are also discussed here.)  In section 4
we derive the diffusion equations, complete with the CP-odd source terms from
the Boltzmann equations, using a truncated expansion in moments of the
distribution functions.  In section 5, these general results are applied to
find and solve the appropriate set of diffusion equations which determine the
chiral quark asymmetry in the MSSM. The rate of baryon production due to the
excess of left-handed quarks is also computed in section 5, and our numerical
results are given in section 6.  In section 7 we present our conclusions, and
a discussion of how the present results differ from previously published
ones.

\section{Introductory example: Fermion with complex mass}

To understand some of the subtleties which arise when solving the equations
of motion in the WKB approximation, let us first consider the example of a
single Dirac fermion with a spatially varying, complex mass,
\beq
        (i\gamma^\mu\partial_\mu - m P_R - m^* P_L)\psi = 0;
        \qquad m = |m(z)| e^{i\theta(z)},
\label{dirac1}
\eeq
where $P_{R,L} = \frac{1}{2}(1 \pm \gamma_5)$ are the chiral projection
operators.  We wish to solve eq.\ (\ref{dirac1}) approximately in an
expansion in gradients of $|m|$ and $\theta$. To simplify the solution we
boost to the frame in which the momentum parallel to the wall is zero
($p_x=p_y=0$) and consider first positive energy eigenstates, $\psi\sim
e^{-i\omega t}$. Then, because spin is a good quantum number, we can write
the spin eigenstate as a direct product of chirality and spin states
\beq
\Psi_s \equiv e^{-i\omega t}
                \left(
                  \begin{array}{cc}
                     L_s \\
                     R_s
                  \end{array}
                \right) \otimes \chi_s ;  \qquad
\sigma_3 \chi_s \equiv s\, \chi_s,
\label{psilambda}
\eeq
where $R_s$ and $L_s$ are the relative amplitudes for right and left
chirality, respectively (and we are using the chiral representation
of the Dirac matrices). Spin $s$ is related to helicity $\lambda$ by
$s \equiv \lambda\, {\rm sign}(p_z)$.  Inserting (\ref{psilambda}) into
the Dirac equation (\ref{dirac1}) then reduces to two coupled complex
equations for complex parameters $L_s$ and $R_s$:
\beqa
\label{dirac2L}
(\omega - is \partial_z)L_s &=& m R_s \\
\label{dirac2R}
(\omega + is \partial_z)R_s &=& m^* L_s.
\eeqa
We can now use eq.\ (\ref{dirac2L}) to eliminate $R_s$
from (\ref{dirac2R}), which then becomes a single second order complex
differential equation for $L_s$:
\beq
\left( (\omega + is \partial_z) \frac{1}{m}
             (\omega - is \partial_z)
            - m^* \right) L_s  = 0.
\label{dirac3}
\eeq
To facilitate the gradient expansion, we write the following WKB ansatz
for $L_s$:
\beq
       L_s \equiv w e^{i\int^z p_c(z') dz'}.
\label{wkbans}
\eeq
We have suppressed the spin index $s$ in $w$ and $p_c$ for simplicity.
Inserting (\ref{wkbans}) into eq.\ (\ref{dirac3}) we find the following
two coupled equations (real and imaginary parts of (\ref{dirac3})):
\beqa
\label{dirac4o}
\omega^2 - |m|^2 - p_c^2 + (s\omega + p_c) \theta' -
        \frac{|m|'}{|m|} \frac{w'}{w} + \frac{w''}{w} &=& 0 \\
\label{dirac4w}
2p_c w' + p_c' w - \frac{|m|'}{|m|}(s\omega + p_c)w - \theta' w' &=& 0
\eeqa
While complicated in appearance, eqs.\ (\ref{dirac4o}-\ref{dirac4w}) are
easily solved
iteratively. For example, to the lowest order one sets all derivative
terms to zero, whereby the first equation immediately gives the usual
dispersion relation $\omega^2 = p_c^2 + |m|^2$. It is also easy to
extend the dispersion relation to first order in derivatives, because
the contributions proportional to $w'$ decouple from eq.\ (\ref{dirac4o})
at this order:
\beq
p_c = p_0 + \scp \frac{s\omega + p_0}{2p_0} \theta' + \alpha',
\label{drsimple}
\eeq
where $p_0 = {\rm sign}(p)\sqrt{\omega^2 - |m|^2}$.  In(\ref{drsimple})
we have shown the
generalization to antiparticles by including the sign $\scp$, which is $+1$
for the particle and $-1$ for the antiparticle.  This follows from the fact
that the Dirac equation for antiparticles is obtained from (\ref{dirac1}) by
the substitution $m \rightarrow -m^*$, which changes $\theta$ to $-\theta$.
The arbitrary function $\alpha'(z)$ reflects the ambiguity in the definition
of momentum $p_c$ in (\ref{wkbans}), because of the freedom to perform
vector-like phase redefinitions of the field, $\psi \rightarrow
e^{i\alpha(z)}\psi$, which cause $p_c \rightarrow p_c + \alpha'$.
This `gauge' dependence reflects the fact that $p_c$ is not the
physical momentum of the WKB-state, a quantity which we will explicitly
identify and show to be gauge independent below. In the preceeding, we
considered the left-handed spinor $L_s$.  The same procedure applied to
$R_s$ gives
\beq
p_c = p_0 + \scp \frac{s\omega - p_0}{2p_0} \theta' + \alpha',
\label{drsimpleR}
\eeq
because of the sign difference between eqs.\ (\ref{dirac2L}) and
(\ref{dirac2R}).  The factors $(s\omega + p_c)$ are likewise
replaced by $s\omega - p_c$ in (\ref{dirac4o}) and
(\ref{dirac4w}).  In the following we will show that this difference
actually does not have any physical effect: the group velocity and
force acting on the particle is the same whether one uses
(\ref{drsimple}) or (\ref{drsimpleR}).   For simplicity we continue
to refer to the relations for $L_s$ unless the contrary is explicitly
stated.

\subsection{Canonical equations of motion}

As anticipated above $p_c$ can be identified as the {\em canonical}
momentum for the motion of the WKB wave-packets. To see this more
clearly, let us first invert (\ref{drsimple}) to obtain an expression
for the invariant energy
\footnote{This discussion is closely analogous to the motion of a particle
in an electromagnetic field, which can be described by a Hamiltonian
\[   H = \sqrt{({\bf p}_c - e{\bf A})^2 + m^2} + e A_0. \]
Here the canonical momentum ${\bf p}_c$ is related to the physical, kinetic
momentum ${\bf p} \equiv m {\bf v}/\sqrt{1-v^2} = \omega {\bf v}_g$ by
the relation ${\bf p}_c = {\bf p} + e{\bf A}$. Canonical momentum is
clearly a gauge dependent, unphysical quantity, because the vector potential
is gauge variant.  Similarly  canonical force acting on ${\bf p}_c$ is gauge
dependent, but the gauge dependent parts cancel when one computes the
physical force acting on kinetic momentum:
\[   \dot {\bf p}_k = -\partial_{\bf x} H - e \partial_t {\bf A}
                       = e({\bf E} + {\bf v}\times {\bf B}).  \]
}
\beq
        \omega = \sqrt{(p_c - \acp)^2 + |m|^2} - \scp \frac{s\theta'}{2},
\label{engy}
\eeq
where $\acp \equiv \alpha' + \scp \theta'/2$ in the left- and $\acp
\equiv \alpha' - \scp \theta'/2$ right chiral sector. (This difference
in $\acp$ has no consequence what follows, which is why we have
suppressed the indices referring to chirality).  Identifying the
velocity of the WKB particle with the group velocity of the wave-packet
(corresponding to the stationary phase condition of the WKB-wave)
it can be computed as
\beqa
       v_g = (\partial_{p_c} \omega )_x
&=&\frac{p_c - \acp}{\sqrt{(p_c - \acp)^2 + |m|^2}}
\nn \\
&=& \frac{p_0}{\omega} \NBR{1 +\scp \frac{s|m|^2\theta'}{2p_0^2\omega}},
\label{vgroup}
\eeqa
where the latter form follows on expanding to linear order in $|m|^2
\theta'/\omega$ after eliminating $p_c -\acp $ with (\ref{engy}). $v_g$
is clearly a physical quantity, independent of the ambiguity in definition
of $p_c$. Given energy conservation  along the trajectory we then have
the equation of motion for the canonical momentum viz.
\beq
\dot{p_c} = -(\partial_x \omega)_{p_c}=v_g \acp'
-\frac{|m||m|' \omega }{(\omega + \scp \frac{s \theta'}{2})}
   + \scp\frac{s\theta''}{2}
\label{ceom-mom}
\eeq
which, like the canonical momentum itself, is
manifestly a gauge dependent quantity, through the first term.
Equations (\ref{vgroup}) and (\ref{ceom-mom}) together are
the canonical equations of motion defining the trajectories of
our WKB particles in phase space.

The physical kinetic momentum can now be defined as corresponding to
the movement of a WKB-state along its world line
\beq
        p \equiv \omega v_g.
\label{phys}
\eeq
This relation also defines the physical dispersion relation between
the energy and kinetic momentum. We now calculate, using the canonical
equations of motion (\ref{vgroup}) and (\ref{ceom-mom}), the force acting
on the particles defined as in eq.\ (\ref{GVF}) i.e. $F= \dot{p} = \omega
\dot{v}_g$, where the latter follows trivially since $\dot{\omega}=0$ along
the particle trajectory. In particular we wish to verify explicitly
that we obtain a gauge independent result for the force.
Using the canonical equations of motion we have
\beqa
     \dot v_g &=&  \dot{x} (\partial_x v_g )_{p_c} + \dot{p_c}
     (\partial_{p_c} v_g)_x  \nonumber \\
       &=&  v_g (\partial_x v_g )_{p_c} - (\partial_x \omega)_{p_c}
     (\partial_{p_c} v_g)_x.
\label{pkdot}
\eeqa
Using the form (\ref{vgroup}) for $v_g$, differentiating and
substituting with the dispersion relation (\ref{engy}), we find
\beqa
(\partial_x v_g )_{p_c} &=&
\frac{m^2}{(\omega + \scp \frac{s \theta'}{2})^{3}}
\nonumber \\
(\partial_{p_c} v_g )_{x}&=&-\acp'
\frac{m^2}{(\omega + \scp \frac{s \theta'}{2})^{3}} -
v_g \frac{|m||m|'}{(\omega + \scp \frac{s \theta'}{2})^2}
\label{rescalings}
\eeqa
from which it is easy to see that the gauge terms (in $\acp$) cancel
out exactly in (\ref{pkdot}) and that the force is given by the gauge
independent expression
\beq
       \dot p = \omega \dot{v}_g=
-\frac{|m||m|' \omega }{(\omega + \scp \frac{s \theta'}{2})^2}
   + \scp\frac{s\theta''}{2}
\frac{|m|^2 \omega }{(\omega + \scp \frac{s \theta'}{2})^3}
\label{force1}
\eeq
which to linear order in $\theta'$ can be written as
\beq
       \dot p = - \frac{|m||m|'}{\omega}
                 + \scp\frac{s(|m|^2\theta')'}{2\omega^2}.
\label{force2}
\eeq
The force therefore contains two pieces. The first is a CP-conserving
part, leading to like deceleration of both particles and antiparticles because
of the increase in the magnitude of the mass. The second part, proportional to
the gradient of the complex phase of the mass term, is CP-violating, and causes
opposite perturbations in particle and antiparticle densities.

In connection with eq.\ (\ref{drsimpleR}) we mentioned the difference in
definition of canonical momentum for left- and right-handed particles.
    From the immediately preceding discussion we can see that this difference
gets absorbed into the definition of the unphysical phase $\acp$. Indeed,
for the right-handed fermions one should define $\acp = \alpha'
-\scp\theta'/2$ instead of $\alpha'+\scp\theta'/2$.  Since we have just
shown that $\acp$ cancels out of physical quantities, the difference
between the dispersion relations derived from the spinors $L_s$ and $R_s$
has no physical effect.  On the other hand, it is true that for
relativistic particles $L_s$ will represent a particle with mostly
negative helicity and $R_s$ will correspond to a mostly positive helicity
particle. The information about helicity ($\lambda$) is contained in the
spin factor, $s = \lambda\, {\rm sign}(p_z)$, and this does have a
physical effect:  particles with opposite spin feel opposite CP-violating
forces.

\subsection{Currents}

Let us conclude this section by considering currents of WKB states
under the influence of the CP-violating classical force (\ref{force2}).
The current can be defined in the usual way,
\beq
j^\mu(x) =  \bar \psi(x) \gamma^\mu \psi(x).
\label{current}
\eeq
Now eq.\ (\ref{dirac4w}) can be used to solve for $w$ to
first order in gradients. After some straightforward algebra one finds
to this order the solution
\beq
\psi_{p,s} = \frac{|m|}{\sqrt{2p^+_s(\omega + sp_0)}}
               \left( \begin{array}{c}
                   1 \\
                   \frac{\omega + sp^+_s}{|m|}
                   \left(1 - \frac{i \lambda \omega|m|'}{2p_0^2|m|}\right)
                      \end{array} \right)  \otimes  \chi_s \;
                  e^{i\int \tilde p_s + i\frac{\theta}{2}\gamma_5 +i\alpha'}
\eeq
where $\tilde p_s \equiv p_0 + s \omega \theta'/(2p_0)$ and $p_s^+ \equiv
\tilde p_s + \theta'/2$. With this expression, it is simple to show
by direct substitution that the current (\ref{current}) corresponding to
a WKB-state becomes
\beq
j^\mu_{p}(x) = \left(\frac{1}{v_{g}} \, ; \,\hat {\bf p}\right),
\label{virta}
\eeq
where we have restored the trivial dependence on ${\bf p}_{||} =
(p_x,p_y,0)$ by boosting in the direction parallel to the bubble wall.
This result confirms the intuitive WKB-particle interpretation; in the
absence of collisions the quasiparticles follow their semiclassical
paths, and when they slow down at some point the outcome is an increase
of local density proportional to the inverse of the velocity.  Because of
this compensation of reduced velocity by increased density, the 3-D
particle flux (${\bf j}$) remains unaffected by the classical force.

Let us finally compute the current arising from a distribution of
WKB-quasiparticle states using the physical dispersion relation. Under our
basic assumption that the plasma can be well described by a collection of
WKB-states we can write
\beq
j^\mu(x) = \int \frac{d^3p\,d\omega}{(2\pi )^4} p^\mu f(\omega ) \,
(2\pi) \,
           \delta \left(\omega^2 - \omega_0^2 +\scp \frac{s|m|^2\theta'}{\omega}
\right).
\eeq
After integrating over $p_z$, this becomes
\beq
j^\mu(x) = \int \frac{p_{||}^2dp_{||}d\omega }{2\pi^2}
             \left(\frac{1}{v_g} \, ; \,\hat {\bf p}\right)
             f(\omega ).
\label{current2}
\eeq
in perfect agreement with the result (\ref{virta}). The current
(\ref{current2}) was recently derived from more fundamental principles
in ref.\ \cite{JKPf} (see also \cite{JKPb}); it was argued that the slightly
more general result obtained in \cite{JKPf} would reduce to the form
(\ref{current2}) in the limit of frequent decohering scatterings; this
limit is of course an underlying assumption in the WKB quasiparticle
approximation used here.

\section{Application of WKB to the MSSM}

In this section we extend the previous analysis of dispersion relations
and canonical equations to the case of the MSSM. The most natural
candidate to effect baryogenesis in the MSSM would appear to be the
left chiral quarks themselves, because any CP-odd perturbations in their
distributions should directly bias the sphaleron interactions. However,
in the MSSM the Higgs field potential is real at tree level, and
therefore the  CP-violating effect on quark masses arises only at one
loop order.  Moreover, the contribution from CP violation present in
the supersymmetric version of the CKM matrix is potentially suppressed by
the GIM mechanism, like in the case of the SM \cite{CPDEB}.
Excluding a direct source in quarks, one must look
for CP-violating sources in various supersymmetric particles. These
species include squarks, which couple to quarks via strong supergauge
interactions, and charginos, which couple strongly to the third family
quarks via Yukawa interactions.  We also comment on neutralinos, which
have couplings similar to those of the charginos.

\subsection{Squarks}

After quarks the natural candidate to consider in the supersymmetric
spectrum are the scalar partners of the third family quarks.  The top
squark mass matrix can be written as
\beq
        M^2_{\tilde q} =
            \left(
               \begin{array}{cc}
                 m^2_Q  & y(A^* H_2 + \mu H_1) \\
                 y(A H_2 + \mu^* H_1) & m^2_U
               \end{array}
            \right)
\label{squarkmass}
\eeq
in the basis of the left- and right-handed fields $\tilde q = (\tilde
t_L$, $\tilde t_R)^T$.  Here the spatially varying VEV's $H_i$ for the
two Higgs fields are normalized such that in the zero temperature vacuum
$\sqrt{2}(H_1^2 + H_2^2) = 246$ GeV. The parameters $m^2_{Q,U}$ refer
to the sum of soft SUSY-breaking masses and VEV-dependent $y^2 m_t^2$ and
$D$-terms, but their explicit form will not be important here. Since
squarks are bosons, they obey the Klein-Gordon equation
\beq
(\partial_t^2 -\partial_z^2 + M^2_{\tilde q})\tilde q = 0.
\label{kgeq}
\eeq
As in the case of a Dirac fermion,  we first boost to the frame where
the particle is moving orthogonal to the wall ($p_x=p_y=0$).  The
chiral structure encountered in the fermionic case is missing here,
but the problem is complicated by left-right flavor mixing.
To deal with this mixing, at first order in the derivative expansion,
it is easiest to perform a unitary rotation $U_q$ to the eigenbasis
of $M^2_{\tilde q}$. The explicit form of the rotation matrix is
\beq
U_q =  {\diag}(e^{i\phi_{qi}})\;
         \frac{\sqrt{2}}{\sqrt{\Lambda_q (\Lambda_q + \Delta_q)}}
        \left( \begin{array}{cc}
           \frac{1}{2}(\Lambda_q+\Delta_q) &           a_q                 \\
                       - a_q^*     & \frac{1}{2}(\Lambda_q+\Delta_q)
                  \end{array} \right),
\label{Uq}
\eeq
where $a_q \equiv y(A_q^*H_2 + \mu^*H_1) $, $\Delta_q \equiv m_Q^2 - m_U^2$,
and $\Lambda \equiv \sqrt{\Delta_q^2 + 4|a_q|^2}$. The diagonal matrix
${\diag}(e^{i\phi_{qi}})$ contains arbitrary phases by which the local
mass eigenstates can be multiplied, or equivalently the ambiguity in the
choice of the rotation matrix, due to the $U(1)$ gauge invariance of the
lagrangian.  After the rotation, eq.\ (\ref{kgeq}) becomes
\beq
\left( \omega^2 + \partial_z^2  - M_d^2  + U_{2}
         + 2U_{1} \partial_z \right) \tilde q_d = 0,
\label{kgerot}
\eeq
where $M^2_d$ is a diagonal matrix, $U_{1} \equiv U_q^\dagger \partial_z
U_q $  and $U_{2} \equiv U_q^\dagger \partial_z^2U_q$. We can formally write
(\ref{kgerot}) as
\beqa
D_{--} \tilde q_- - D_{-+} \tilde q_+ &=& 0 \nn \\
D_{++} \tilde q_+ - D_{+-} \tilde q_- &=& 0.
\label{kgeq2}
\eeqa
The quantities $D_{\pm\mp}$'s are differential operators in the rotated
basis, which makes it impossible to exactly decouple the equations for
the variables $\tilde q_\pm$.  However, one can show that they do
decouple to first order in the gradient expansion. To this
end we first write the equations (\ref{kgeq2}) in the form
\beqa
(D_{++} D_{--} - D_{-+}D_{+-}) \tilde q_-
             + [D_{-+},D_{++}] \tilde q_+ &=& 0
\nn \\
(D_{--} D_{++} - D_{+-}D_{-+}) \tilde q_+
             + [D_{+-},D_{--}] \tilde q_- &=& 0.
\label{kgeq3}
\eeqa
It is easy to see that the commutator terms are of second order or higher
in derivatives of mass matrix elements.  Similarly, the products of the
off-diagonal terms $D_{\pm\mp}D_{\mp\pm}$ are second order or higher and
can be neglected.  Finally, one can show that $D_{\mp\mp}D_{\pm\pm}\tilde
q_\pm = c_\pm \, D_\pm \tilde q_\pm + {\cal O}(\partial_z^2)$, where
$c_\pm$ are some constants.  One then has simply
\beq
D_{\pm\pm} \tilde q_\pm = 0
\label{kgeq4}
\eeq
up to second order gradient corrections. Inserting the WKB ansatz
$\tilde q_\pm \equiv w_\pm e^{i \int^z p_{c\pm} dz}$ into (\ref{kgeq4})
one finds
\beq
       \left (2 ip_{c\pm} w_\pm' + ip_{c\pm}' + \omega^2 - m^2_\pm
       - p^2_{c\pm}
       + 2i p_{c\pm} U_{1\pm\pm}\right) w_\pm = 0.
\label{cpldsquark}
\eeq
Breaking up the real and complex parts of the equations, we get
\beqa
      \omega^2 - m^2_\pm - p^2_{c\pm} &=& 2 p_{c\pm}{\rm Im}(U_{1\pm\pm}),
\label{cpldsquark2a} \\
       2 p_{c\pm} \frac{w_\pm'}{w_\pm}
                         + ip_{c\pm}' &=& 2 p_{c\pm}{\rm Re}(U_{1\pm\pm}).
\label{cpldsquark2b}
\eeqa
The correction term $U_{1\pm\pm}$ appearing in the above equations is
in fact purely imaginary:
\beqa
U_{1\pm\pm}
                &\equiv& i \theta_{q\pm}'
\nn \\
                &=& \mp \frac{2iy^2}{\Lambda_q(\Lambda_q+\Delta_q)}
                   {\rm Im}(A_t^*\mu) (H_1'H_2 - H_2'H_1)
                   + i \tilde \phi_{q\pm}' \,,
\label{thetaq}
\eeqa
where $\tilde \phi_{q\pm}' = (U_q {\diag}(\phi_{qi}')
U_q^\dagger)_{\pm\pm}$ are still some arbitrary phases.
Using this notation, we see that the dispersion relation acquires
an energy-independent shift from the leading order result:
\beq
p_{c\pm} = p_{0\pm} - \theta_{q\pm}',
\label{DRsq}
\eeq
where $p_{0\pm} \equiv {\rm sign}(p_z) \sqrt{\omega^2 - m_\pm^2}$.
Curiously, the parametric form of the shift (\ref{thetaq}) is the same
as what appears in the source derived for squarks in \cite{Riotto}. In
our method this correction originated from a local rotation in the flavour
basis of the mass eigenstates. Similarly, in \cite{Riotto} the source
$\propto {\rm Im}(A_t^*\mu) (H_1'H_2 - H_2'H_1)$ was found
by performing an expansion to a finite order in the flavour nondiagonal
mass insertion over temperature, which is an approximate way of taking
into account a rotation of eigenstates in a varying background. In the
present context we can see, however, that this shift is unphysical,
because of
the arbitrariness of the phases $\tilde \phi_{q\pm}'$ in (\ref{thetaq}).
For example, they could be chosen to make  $\theta_{q\pm}'=0$.  This is
only possible because the expression  (\ref{thetaq}) is a function of
$x$ only, and not $p$. Indeed, proceeding in analogy with the fermionic
case of section (2.1), we find that $p_{c\pm}$ is to be identified as
the canonical momentum of the system. Moreover, defining the physical
momentum $p_\pm$ through the group velocity as in (\ref{phys}), we find
\beq
p_\pm  \equiv  \omega v_{g\pm} = p_{0\pm}.
\eeq
Thus the physical momentum gets no corrections to first order
accuracy.  This implies that neither does any classical force arise  at
first order in gradients.  Notice also that, because ${\rm
Re}(U_{1\pm\pm})=0$, the normalization of the state can be  computed
only to the zeroth order from (\ref{cpldsquark2b}), which gives $w_\pm
= C_\pm/\sqrt{p_{0\pm}}$, where $C_\pm$ are constants.

Because the CP-violating source can only arise at second order in the
gradient expansion in the squark sector, it is parametrically small
compared to a fermionic source (to be derived for charginos below). Given
the range of wall widths compatible with a sufficiently strongly first order
phase transition in the MSSM \cite{MQS,CM}, we can estimate this suppression
roughly to be of the order $\sim k m'/m \sim 1/3T\ell_w \sim 1/30$ for a
particle with thermal de Broglie wave number $k\sim 1/3T$ and wall width
$\ell_w\sim 10/T$. We will accordingly neglect the squark source henceforth.

\subsection{Charginos}

An asymmetry in Higgsinos is efficiently transported to left-handed quarks
via strong Yukawa interactions with third family quarks. The chargino mass
term,
\beq
\label{cmt}
	\overline\Psi_R M \Psi_L + {\rm h.c.},
\eeq
contains complex phases required for a CP-violating force term. In the
basis of Winos and Higgsinos the chiral fields are
\beqa
	\Psi_R &=& (\tilde W^+_R, \tilde h^+_{1,R})^T \nonumber\\
\label{cmtfields}
	\Psi_L &=& (\tilde W^+_L, \tilde h^+_{2,L})^T
\eeqa
and the mass matrix is
\beq
       M = \left(
                  \begin{array}{cc}
                      m_2  &  g H_2 \\
                     g H_1 &   \mu
                  \end{array}
                \right),
\label{charginomass}
\eeq
where the spatially varying VEV's $H_i$ are definded as in the squark
case. The corresponding Dirac equation, in the frame where $p_x=p_y=0$,
is
\beq
        ( i\gamma_0 \partial_t - i\gamma_3 \partial_z
         -  M^\dagger P_L - M P_R) \Psi = 0.
\label{chDirac1}
\eeq
To solve it in the WKB approximation, we follow the same procedure as with
the single Dirac fermion and the squark cases above.  First we introduce
the spin eigenstate as a direct product of chirality and spin states,
where spin $s$ and helicity $\lambda$ are related by $s \equiv \lambda\,
{\rm sign}(p_z)$:
\beq
\Psi_s \equiv e^{-i\omega t}
                \left(
                  \begin{array}{cc}
                     L_s \\
                     R_s
                  \end{array}
                \right)\otimes \Phi_s ;  \qquad
\sigma_3\Phi_s \equiv s\, \Phi_s.
\label{psilambda2}
\eeq
In contrast with the simple Dirac fermion, the relative amplitudes of
left and right chirality, $L_s$ and $R_s$, are now two-dimensional
complex vectors in the Wino-Higgsino flavor space. Keeping this
generalization in mind, the solution proceeds formally in analogy to
the case of a single Dirac fermion; inserting (\ref{psilambda2})
into the Dirac equation (\ref{chDirac1}) gives
\beqa
\label{chDirac2L}
        (\omega - is\,\partial_z)L_s &=& M R_s \\
\label{chDirac2R}
        (\omega + is\,\partial_z)R_s &=& M^\dagger L_s.
\eeqa
    From eq.\ (\ref{chDirac2L}) we have $R_s = M^{-1}(\omega - is \pz ) L_s$,
which when substituted into (\ref{chDirac2R}) gives
\beq
\left( \omega^2 + \pz^2 - M M^\dagger
            + is(M \pz M^{-1})(\omega -is\pz ) \right) L_s = 0.
\label{chDirac3}
\eeq
Since $L_s$ is a two-component object, writing the WKB-ansatz is
somewhat more involved than it is for a single fermion. But since
$M M^\dagger$ is a hermitian matrix, we can rotate to its diagonal
basis, similarly to the squarks.  Eq.\ (\ref{chDirac3}) then becomes
\beqa
\left( \omega^2 + \pz^2 - m_D^2 + 2\Uso\pz + \Ust
                 + is \Aso(\omega -is\pz ) + \Ast \right) L^d_s = 0,
\label{chDirac4}
\eeqa
where the superscript in $L^d_s$ indicates that we are in the basis
where $M M^\dagger$ is locally diagonal as a function of distance
$z$ from the wall. The $2\times 2$ matrices (in the Wino-Higgsino flavor
space) in (\ref{chDirac4}) are defined by
\beqa
        && \Uso \equiv U\pz U^\dagger; \qquad  \phantom{Hannati}
           \Ust \equiv U\pz^2 U^\dagger;  \phantom{Hann} \nn \\
        && \Aso \equiv U (M\pz M^{-1}) U^\dagger;  \qquad
           \Ast \equiv \Aso \Uso. \phantom{Hann}
\label{matrices}
\eeqa
The explicit form of the rotation matrix $U$ which diagonalizes
$M M^\dagger$ is similar to the one encountered in the squark case:
\beq
U =  {\diag}(e^{i\phi_i})\;
         \frac{\sqrt{2}}{\sqrt{\Lambda (\Lambda + \Delta)}}
        \left( \begin{array}{cc}
             \frac{1}{2}(\Lambda+\Delta) &           a                 \\
                          - a^*          & \frac{1}{2}(\Lambda+\Delta)
                  \end{array} \right),
\label{U}
\eeq
where
\beqa
         a     &\equiv&  m_2u_1 + \mu^*u_2 \nn \\
       \Delta  &\equiv& |m_2|^2 - |\mu |^2 + u_2^2  - u_1^2 \nn \\
       \Lambda &\equiv& \sqrt{\Delta^2 + 4|a|^2}\nn\\
	   u_i &\equiv&gH_i.
\eeqa
The arbitrary angles $\phi_i$ will eventually enter the dispersion relation
as physically irrelevant shifts in the canonical momenta, similarly to the
squark case and the case of the single Dirac fermion. The diagonalized
$M M^\dagger$ matrix, $m_D^2 = {\diag}(m_+^2,m_-^2)$, has the
eigenvalues
\beq
      m_\pm^2  =  \frac{1}{2}\left(|m_2|^2 + |\mu |^2 + u_2^2  + u_1^2\right)
                   \pm \frac{\Lambda}{2}.
\label{masses}
\eeq

Broken into components, labeled by $\pm$, and suppressing the spin
index on $L^d_s$, equation (\ref{chDirac4}) can be written as
\beqa
D_{--} L^d_- - D_{-+} L^d_+ &=& 0 \nn \\
D_{++} L^d_+ - D_{+-} L^d_- &=& 0.
\label{chDirac5}
\eeqa
Just as in the squark case, one can show that the mixing terms in
(\ref{chDirac5}) can be neglected to the first order in the gradient
expansion, and it is sufficient to solve the decoupled equations
\beq
        D_{\mp\mp} L^d_\mp = 0.
\label{chDirac6}
\eeq
Inserting the WKB ansatz, $L^d_\pm \equiv w_\pm e^{i\int p_\pm dz}$, into
(\ref{chDirac6}), and writing $D_{\mp\mp}$ explicitly, we obtain
\beq
\left( \omega^2 - p_\pm^2 - m_\pm^2 + ip_\pm' + 2ip_\pm\pz + 2ip_\pm
\Uso_{\pm\pm}
+ is(\omega + sp_\pm)\Aso_{\pm\pm} \right)\, w_\pm = 0.
\label{1order1}
\eeq
Taking the real and imaginary parts of this equation we have
\beqa
\omega^2 - p_\pm^2 - m_\pm^2  &=&
      \Im \left( 2p_\pm \Uso_{\pm\pm} + s(\omega + sp_\pm)\Aso_{\pm\pm} \right)
\label{eveqs1}
\\
p_\pm' + 2p_\pm \frac{w_\pm'}{w_\pm} &=&
      \Re \left( 2p_\pm \Uso_{\pm\pm} + s(\omega + sp_\pm)\Aso_{\pm\pm} \right).
\label{eveqs2}
\eeqa
Equations (\ref{eveqs1}-\ref{eveqs2}) are similar to the equations
(\ref{cpldsquark2a}-\ref{cpldsquark2b}) for squarks apart from the
appearance of new contributions from the matrix $A_1$ in (\ref{eveqs1}-
\ref{eveqs2}); as we shall see, this difference is crucial.
Eq.\ (\ref{eveqs1}) gives the dispersion relation to first order in
gradients; however (\ref{eveqs2}) gives the normalization ($w_\pm$) only
to zeroth order, because integrating $w'_\pm$ eliminates one derivative.
To this order $w_\pm$ give the usual spinor normalization, but with a
spatially varying mass terms.  If we needed to know $w_\pm$ also at first
order, it would be necessary to include second order corrections to
(\ref{eveqs2}). Luckily we do not need these results here, and therefore
concentrate on the dispersion relation in the following.

To find the leading correction to the dispersion relation we need to
compute the diagonal elements of the matrices $\Uso$ and $\Aso$:
\beqa
\Im \Aso_{\pm\pm} &=& \pm \frac{\Im (m_2\mu )}{m^2_\pm \Lambda}
                         (u_1u_2' + u_2u_1')  \nn \\
\Im \Uso_{\pm\pm} &=& \pm \frac{2 \Im (m_2\mu )}{\Lambda (\Lambda +\Delta)}
                         (u_1u_2'-u_2u_1') + i {\tilde \phi}_{L\pm}',
\label{UetAdiag}
\eeqa
where
${\tilde \phi}_{L\pm}' \equiv (U {\diag}(\phi_i') U^\dagger)_{\pm\pm}$.
Defining $p_0 = {\rm sign}(p)\sqrt{\omega^2 - |m_\pm|^2}$ the dispersion
relation for the states associated with $L^d_{\pm}$ becomes
\beqa
        p^L_\pm \; = \; p_{0\pm}
               &\mp& \scp \frac{s(\omega + sp_{0\pm} )}{2p_{0\pm}}
                   \frac{\Im (m_2\mu )}{m^2_\pm \Lambda} (u_1u_2' + u_2u_1')
\nn \\
               &\mp& \scp \frac{2 \Im (m_2\mu )}{\Lambda(\Lambda +\Delta)}
                   (u_1u_2'-u_2u_1')
                  \; \pm \; i \Lambda \tilde \phi_{L\pm}'.
\label{charginoDR}
\eeqa
The sign $\scp$ is 1 ($-1$) for particles (antiparticles).
The signs $\pm$ refer to the mass eigenstates; below, we will want to
focus on the state which smoothly evolves into a pure Higgsino in
the unbroken phase in front of the wall.
This will depend on the
hierarchy of the diagonal terms in the chargino mass matrix  in the
following way:
\beq
\label{whichp}
	p_{\tilde{h}_2} = \left\{\begin{array}{ll} p^L_+, & |\mu| > |m_2| \\
					p^L_-, & |\mu| < |m_2|.
	\end{array} \right.
\eeq
The reason for identifying $p$ with the Higgsino $\tilde{h}_2$ is that
it plays the role of the left-handed species in the mass term, as we have
written it in eqs.\ (\ref{cmt}-\ref{cmtfields}).

In the diffusion equations to be derived in the following
sections, we will treat the charginos as relativistic particles, whose
chirality is approximately conserved.  We therefore would also like to
know the dispersion relation for the other flavor component,
$\tilde{h}_2$, which is associated with the right-handed spinor
$R^d_s$.  Our convention for the sign of $\tilde{h}_2$-number is the
supersymmetric one, where the Higgsino mass term is written in terms of
left-handed fields only and has the form
\beq
	\mu \tilde h_{1} \tilde h_{2} + {\rm h.c.}
\eeq
Explicitly, we identify
\beqa
	\tilde h_{1} &\leftrightarrow& \tilde h^+_{1,L}
	                   = (\tilde h^+_{1,R})^c \nonumber\\
	\tilde h_2 &\leftrightarrow& \tilde h^-_{2,L}
\eeqa
That is, $\tilde h_1$ is identified with the CP conjugate of $\tilde
h^+_{1,R}$, whereas $\tilde h^+_{1,R}$ itself is the particle
represented by the spinor $R^d_s$.  Therefore we must remember to
perform a CP conjugation of the $R^d_s$-field dispersion relation if it
is to represent the states which we call $\tilde h_1$, a point which
can be somewhat confusing.\footnote{We erred on this point in \cite{CJK}.}

Going through the same steps as for $L^d_s$ to find the dispersion
relation for $R^d_s$, we obtain
\beq
       \omega^2 - p_\pm^2 - m_\pm^2  =
        \Im \left( 2p_\pm {V_1}_{\pm\pm} + s(\omega - 
sp_k){B_1}_{\pm\pm} \right)
\label{drforh2}
\eeq
where $V_1$ and $B_1$ respectively can be obtained from
$\Uso$ and $\Aso$ by exchanging $u_1
\leftrightarrow u_2$ and taking complex conjugates of $m_2$ and $\mu$.
Taking into account the additional sign change $\scp\to-\scp$ required
by our convention for the meaning of $\tilde h_1$, we obtain
\beqa
        p^{R}_\pm \; = \; p_{0\pm}
            &\mp& \scp \frac{s(\omega - sp_{0\pm} )}{2p_{0\pm}}
                   \frac{\Im (m_2\mu )}{m^2_\pm \Lambda} (u_1u_2' + u_2u_1')
\nn \\
            &\pm& \scp \frac{2 \Im (m_2\mu )}{\Lambda(\Lambda + \bar\Delta)}
                   (u_1u_2'-u_2u_1') \mp i \Lambda \tilde \phi_{R\pm}'.
\label{charginoDR2}
\eeqa
where $\bar \Delta \equiv |m_2|^2 - |\mu |^2 + u_1^2  - u_2^2$,
${\tilde \phi}_{R\pm}' \equiv (V {\diag}(\phi_i') V^\dagger)_{\pm\pm}$,
and by definition $\scp=1$ for the state $\tilde h_{1,L}$. Similarly to
(\ref{whichp}), we identify
\beq
\label{whichp2}
	p_{\tilde{h}_1} = \left\{\begin{array}{ll} p^R_+, & |\mu| > |m_2| \\
					p^R_-, & |\mu| < |m_2|.
	\end{array} \right.
\eeq

The term proportional to $u_1u_2'-u_2u_1'$ in (\ref{charginoDR}) and
(\ref{charginoDR2}) was not included in our earlier work. (The complete
dispersion relation was however given in reference \cite{KK} recently.)
This term is odd under exchange of the higgs fields, and appears potentially
viable to produce a source $S_{H_1-H_2}$ in the diffusion equations
for the combination $\tilde h_{1} - \tilde h_{2}$. Indeed, if one
does not keep in mind that the momenta $p^{\sss L,R}_\pm$
in (\ref{charginoDR}) and (\ref{charginoDR2}) are the canonical
momenta, and not the physical momenta, one is easily led to infer
(as recently in ref.\ \cite{HS}) that there is a CP-violating force,
since the canonical equation of motion $\dot{p_c}=-(\partial_x
\omega)_{p_c}$ includes a contribution from this $u_1u_2'-u_2u_1'$ piece.
As discussed in section 2, however,
the latter quantity is, like the canonical momentum itself, a gauge
invariant quantity which changes under arbitrary overall local phase
transformations on the fields. Just as in the squark case, the
$u_1u_2'-u_2u_1'$-part of the dispersion relation is energy-independent,
can be absorbed into the arbitrary phase factor $\Lambda \tilde \phi_\pm'$,
and as such does not
represent a physical quantity. In fact, grouping all unphysical constant
terms from the r.h.s.\ of (\ref{charginoDR}) or (\ref{charginoDR2}) into
a common arbitrary phase factor, we get
\beq
p^{\sss L,R}_\pm \; = \; p_{0\pm}
                    + \scp \frac{s\omega \theta_\pm'}{2p_{0\pm}}
                    + \alpha^{\sss L,R}_\pm,
\label{charginoDR3}
\eeq
with the physical phase $\theta_\pm'$ defined by
\beq
        \theta_\pm' \equiv \mp \frac{\Im (m_2\mu )}{m^2_\pm \Lambda}
                          (u_1u_2' + u_2u_1').
\label{thetaprime}
\eeq
This result shows how the chiral force depends on having both
CP-violating couplings in the Lagrangian (in this case the phase of
$m_2\,\mu$) and spatially varying VEVs ---otherwise the phases could be
removed by global field redefinitions ---so that the force is operative
only within the wall. Treating the whole problem in the kinetic
variables, as we do here, one avoids by construction the problems of
gauge variance one encounters when using the canonical variables.

The physical part of the dispersion relation is identical for both
species of higgsinos, $\tilde h_1$ and $\tilde h_2$.
Hence these states will have the same
group velocities and experience the same physical force in the region of the
wall. The form (\ref{charginoDR3}) for the dispersion relation is also
identical to that for the single Dirac fermion, with the simple replacement
$\theta \rightarrow \theta_\pm$, so that we immediately have
\beqa
v_{g\pm}  &=& \frac{p_{0\pm}}{\omega} +
           \frac{s\, \scp\, m^2_\pm \theta_\pm'}{2\omega^2p_{0\pm}}
\label{vgroupch}
\\
F_{\pm}   &=&  - \frac{m_\pm m_\pm'}{\omega} + \frac{s\,\scp}{2\omega^2}
\left( m_\pm^2 \theta_\pm' \right)'.
\label{chforce}
\eeqa
Since the force terms are identical for both kinds of higgsinos, so
will be the source terms in their respective diffusion equations. This will
become explicit when we prove in the next section that the source is
proportional to a weighted thermal average of the force term. The outcome
is that the linear combination $S_{H_1-H_2}$ considered in \cite{HS} and
in \cite{HN,CQRVW} is not sourced at all in the classical force mechanism,
at leading order in the WKB expansion.

\subsection{Neutralinos}

Neutralinos are an obvious candidate to study after charginos. The mass
term for neutralinos can be  written analogously to that of the charginos
as $\overline\Psi_R M_n \Psi_L +$ h.c., where in the basis of gauginos
and neutral Higgsinos, $\Psi_R =  (\tilde B, \tilde W_3, \tilde h^0_{1,R},
\tilde h^0_{2,R})^T$,
\beqa
              M_n &=& \left(\begin{array}{cc} A & v \\
                              v^T & B \end{array}\right);\qquad
              A = \left(\begin{array}{cc} m_1 & 0\\ 0 & m_2
                      \end{array}\right);\qquad B = \left(\begin{array}{cc}
                              0 & \mu \\ \mu & 0 \end{array}\right);\nonumber\\
              v &=& m_Z\left(\begin{array}{cc} \cos\theta_w\sin\beta &
	-\cos\theta_w\cos\beta \\ -\sin\theta_w\sin\beta &
	\sin\theta_w\cos\beta\end{array}\right)
\eeqa
where $\tan\beta=v_2/v_1$, and $m_Z$ is the Higgs-field-dependent $Z$
boson mass. Because it is a $4\times 4$ matrix it is more difficult to
solve for the WKB eigenstates of the neutralinos than for the
charginos.  However the structure of the mass matrices is sufficiently
similar to suggest that the chiral force on neutralinos and hence the
magnitude of the produced asymmetry in neutralinos is not in any way
parametrically different from that of the charginos, and should be
quantitatively similar as well.
However, the transport of the asymmetry from neutralinos to the quark
sector is much less efficient due to smaller gauge-strength coupling
to fermions. We will therefore limit ourselves to a computation
of the chargino contributions alone in the following estimate of the
baryon asymmetry.

\section {Transport Equations}

Our next task is to determine how the nontrivial dispersion relations
lead to CP-odd perturbations on and around the bubble wall. In
particular we need to determine how an asymmetry in left-handed quarks
is produced which drives the electroweak sphalerons to generate the
baryon asymmetry.  Indeed, our primary source particles with direct
CP-violating interactions with the wall (charginos) experience no
baryon number violating interactions, and therefore the CP-violating
effects must be communicated to the left-handed quark sector via
interactions.  Within the WKB approximation the plasma is described by
a set of Boltzmann equations for the quasiparticle distribution functons.
We will not attempt to solve the full momentum dependent equations;
instead we will use them as a starting point to derive, by means of
a truncated moment expansion, a set of diffusion equations for the
local chemical potentials of the relevant particle species.

The advancing phase transition front (bubble wall) distorts the plasma
away from the equilibrium distributions which would exist if the wall
were stationary.  The exact form of the distortion is complicated, but
a simple ansatz can be made in the present situation for two reasons.
First, the perturbations in the chemical compositions  are small in
amplitude, because they are suppressed by the presumably small phase
in the CP violating part of the force exerted by the wall on the
particles. Second, the elastic interactions enforcing the kinetic
equlibrium are much faster than the ones bringing about the chemical
equilibrium and moreover, they are also very fast compared to the wall
passage time-scale.

We first need to determine the form of the local equilibrium distribution
function for the WKB-states. To this end we need to more accurately specify
what we mean by the particle interpretation of WKB-solutions, \ie what is the
appropriate local $z$-component of the momentum.  Following our reasoning in
the treatment of the flow term, we argue that for an interaction with a mean
free path less than the wall width, $\Gamma^{-1} << \ell_w$, the WKB-state
can be approximated by a plane wave (particle) solution
\beq
    \psi_{\rm WKB} \ra \exp [i(\omega t - {\bf p}_{||} \cdot
                                {\bf x}_{||} - p_z z)],
\label{local}
\eeq
where ${\bf p}_{||}$ is the momentum parallel to the wall and $p_z$ is
the kinetic momentum defined by the physical dispersion relation such as
(\ref{vgroup}) or (\ref{vgroupch}). If $\Gamma^{-1} << \ell_w$ one can to a
good approximation compute the collision terms by extending this wave
solution to infinity, which then formally implies the usual Feynman rules
for the states with conserved 4-momentum $(\omega; {\bf p_{||}}, p_z)$.
Under these considerations the appropriate equilibrium solution in the
wall frame is given by
\beq
      f_{i0}(p,x) = {1\over e^{\beta\gamma_w(\omega + v_w p_z)} \pm 1},
\label{disteq}
\eeq
where $v_w$ is the velocity of the bubble wall, $\gamma_w=1/
\sqrt{1-v_w^2}$, $\beta \equiv 1/T$ and $p$ is the kinetic momentum.
The signs $\pm$ refer to fermions or bosons, respectively. It is clear
that the distribution (\ref{disteq}) makes the collision integral vanish
for the states (\ref{local}) with a constant $p_z$.
However, if $\Gamma^{-1}$ is not small in comparison with the wall width,
one should expect corrections of the order $\sim \Gamma \partial_z$ to the
collision terms. These corrections are parametrically of the form of the
spontaneous baryogenesis terms discussed earlier in the literature
\cite{CKNsp,CKN,JPT,CJK}. We do not attempt to derive these terms here,
because based on our earlier estimates \cite{CJK}, we expect them to be
subleading to the source term deriving from the classical force in the
flow term.

The need to define the equilibrium distribution externally is an inherent
shortcoming of the semiclassical method. In previous papers \cite{JPT,CJK}
a slightly different approach was taken, where the ansatz was parametrized
in terms of the canonical momentum. In canonical parametrization however,
one is faced with a more subtle task to obtain phase reparametrization
invariant physical results. We will give a detailed comparision of the
two approaches in appendix B; here we comment that our present definition
has a more direct physical interpretation and that the difference between
the two approaches would lead to only small changes in our final results.
At any rate, a more complete derivation of the transport equations, using
the methods developed in \cite{JKPf,JKPb}, will be needed to settle the
issue self-consistently. We now extend the ansatz (\ref{disteq}) by allowing
perturbations around the equilibrium:
\beq
      f_i(p,x) = {1\over e^{\beta[ \gamma_w(\omega + v_w p_z) - \mu_i(x)]}
                  \pm 1}  + \delta\! f_i(p,x),
\label{dist}
\eeq
where the local chemical potential function $\mu_i(x)$ and the local
momentum-dependent deviation function $\delta\! f_i(p,x)$ are expected
to be small in magnitude.  The first part of (\ref{dist}) describes the
system in kinetic, but possibly out of chemical equilibrium, whereas the
momentum-dependent part $\delta\! f_i(p,x)$ describes any deviations from
kinetic equilibrium. By definition of the local chemical potential
$\mu_i(x)$, the total number density is given by the first term of
(\ref{dist}) alone, and hence
\beq
\int \rd^3p \; \delta\! f_i(p,x) = 0.
\label{delfco}
\eeq
The $\delta\! f(x,p)$ term will play the role of an auxiliary field in
the following derivation. It cannot be neglected, for the Boltzmann
equations with a force term can only be self-consistently solved if
there is a perturbation which is anisotropic in momentum
space.\footnote{The same is true of the derivation of the diffusion
current $\vec{j}= -D \vec\nabla n$ in the simplest free diffusion
approximation, or Ohm's Law in the presence of an electric potential.}
The shape of $\delta\! f_i(p,x)$ cannot be consistently restricted beyond
(\ref{delfco}), since this would require some hierarchy  between the
elastic interaction rates that would allow certain perturbations to be
damped more slowly than others.

We now return to our derivation of the transport equations, adopting the
ansatz (\ref{dist}). We have derived the dispersion relation in the rest
frame of the bubble wall, where it is also easiest to derive the flow term
of the Boltzmann equation. The collision term is simple in the rest frame
of the plasma however, and to exploit this we  must Lorentz-transform the
flow term to the plasma frame in the end. Since the wall velocities are
expected to be less than 0.1 \cite{Moore,JS}, we will expand in $v_w$
and keep terms to first order in $v_w$; hence $\gamma_w \cong 1$ and
it is sufficient to perform a  Galilean transformation for the flow
term.

In the following subsection we will derive the diffusion equations in
detail. The result will be a set of coupled equations for $\xi_i \equiv
\mu_i/T$ of the form $-D_i\xi_i'' - v_w \xi_i' + \Gamma^d_{ij}\xi_j =
S_i$, where $D_i$ is some yet to be determined diffusion coefficient,
$\Gamma^d_{ij}$ is a matrix of rates at which particles are lost due to
decay and inelastic collision processes, and $S_i$ is the source term
due to the chiral force. Application of this formalism  to the MSSM
will be considered in the next section.


\subsection{Derivation of diffusion equations}

{\bf Moment expansion}. Our starting point is a set of coupled Boltzmann
equations for the relevant particle species. Since the expansion of the
universe is  negligible compared to the diffusion and interaction time
scales, we have simply
\beq
       (\partial_t + {\bf v}_g \cdot \partial_{\bf x}  +
       {\bf F} \cdot \partial_{\bf p}) f_i = C[f_i,f_j,...].
       \label{boltzeq},
\eeq
where ${\bf v}_g$ is the group velocity and ${\bf F}={\dot {\bf p}}$
is the classical force, as derived in sections 2 and 3, and the partial
derivative with respect to ${\bf x}$ is now taken at fixed kinetic
momentum ${\bf p}$.
\footnote{The Boltzmann equation was earlier \cite{JPT,CJK} written
in terms of canonical variables:
\beq
       (\partial_t + {\bf v}_g \cdot \partial_{\bf x}  +
        \dot {\bf p}_c \cdot \partial_{\bf {p_c}}) f_i = C[f_i,f_j,...].
       \label{boltzeq-can},
\eeq
Equations (\ref{boltzeq}) and (\ref{boltzeq-can}) are actually equivalent,
since Eqn.\ (\ref{boltzeq}) can be obtained from (\ref{boltzeq-can}) by a
simple change of variables $(x,p_c) \rightarrow (x,p(p_c,x))$.}
To derive the diffusion equation for $\mu_i$, we
first insert the ansatz (\ref{dist}) into (\ref{boltzeq}),
approximating $\gamma_w = 1$, which gives
\beq
	v_g \BBR{ \nBR{(\partial_z\omega )_{p_z} - \mu'}
	          {\partial f_i\over\partial\omega }
	        + \partial_z \delta f_i }
	+ F_{i,z} \BBR{ \nBR{(\partial_{p_z}\omega )_z + v_w}
	                {\partial f_i\over\partial\omega}
	                    + \partial_{p_z} \delta\! f_i }
       = C[f_i,f_j,...].
\label{step}
\eeq
where primes mean ${\partial/\partial z}$ and the expression
$\partial f_i/\partial\omega$ refers only to the non-$\delta\! f_i$ part
of $f_i$. It should be noted that the partial derivatives with respect to
$p_z$ in (\ref{step}) are taken with respect to the kinetic momentum,
and therefore do not correspond to the usual canonical equations of motion,
\ie $(\partial_{p_z}\omega )_z \neq v_g$ and $-(\partial_z\omega )_{p_z}
\neq F_{i,z}$. However, these terms still cancel because their sum is a
total derivative of $\omega$:
\beq
v_g (\partial_z\omega )_{p_z} + F_{i,z} (\partial_{p_z}\omega )_z
\;= \; \dot z (\partial_z\omega )_{p_z}
          + \dot p_z (\partial_{p_z}\omega )_z
\;=\; 
    {\rm d}\omega/{\rm d}t \;=\; 0.
\label{cancel}
\eeq
Moreover, because the correction $\delta\! f$ is generated by the force,
one can easily see that the CP-even correction arising from the $F_{i,z}
\partial_{p_z} \delta\! f_i$ term is of second order in gradients,
and the CP-odd correction of third order, both one order higher than
corresponding terms coming from the leading $v_w F_{i,z}$ term. We can
thus neglect it and we are left with
\beq
      {\partial f_i\over\partial\omega }
          \left(v_w F_{i,z} - \mu'{p_z\over\omega} \right)
            + \frac{p_{z}}{\omega} \delta\! f_i'
                =   C[f_i,f_j,...],
\label{boltzeq2}
\eeq
where we also wrote $v_g = p_z/\omega$, which is an exact relation in
the physical variables. As expected, the only source term\footnote{\ie
the term which is nonvanishing when $\mu= \delta\! f=0$. Notice that the
collision term vanishes in this limit.}
in (\ref{boltzeq2}) is proportional to the wall velocity, so that
the trivial unperturbed Fermi-Dirac or Bose-Einstein distribution,
$f_0 = (e^{\beta \omega} \pm 1)^{-1}$ becomes a solution in the limit of
a wall at rest in the plasma frame, as it should.  We now wish to transform
to the plasma frame, since the collision integral takes a simple form there.
Since the wall velocity is small \cite{Moore,JS}, we can make a Galilean
transformation of (\ref{boltzeq2}), $v_g \to v_g + v_w$.  We can also
replace $f_i$ by the unperturbed distribution $f_0$ to leading order
in the perturbation.
Then
\beq
        \left( \frac{p_z}{\omega} + v_w\right)
          \left( -\mu_i'\,\frac{\partial f_0}{\partial \omega}  + \delta
f_i' \right)
         +v_w\, \frac{\partial f_0}{\partial \omega} F_{iz}
         = C[\mu_i,\delta\! f_i,...]
\label{boltz2}
\eeq
for the Boltzmann equation (\ref{boltzeq}) in the plasma rest frame.
The term in (\ref{boltz2}) containing the force $F_i$ is what pushes
the distributions out of both kinetic and chemical equilibrium. Notice
that the chemical potential in this equation is not a purely CP-odd
quantity. It is actually a {\em pseudo}chemical potential \cite{DK}
which also has a component caused by the CP-even part of the classical
force.  The latter is relevant for the dynamics of the wall expansion,
and determines the wall velocity and shape \cite{MP,Moore,JS}.  For
baryogenesis however, only the CP-violating part is important. We will
therefore neglect the CP-even parts of $\mu$ and $F_i$ in what follows,
denoting by $\delta\! F_{iz}$ the CP-odd part of the force.

We next integrate both sides of (\ref{boltz2}) over momentum, weighting
by $1$ and by $p_{z}/\omega$, respectively, to obtain two independent moment
equations:\footnote{This choice is not unique. To get the most reliable
truncated equations, the moments should be chosen such that they
sensitively probe the momentum region that is most relevant for the
problem. Here the physical effect is not confined to any sharply cut
region in the momentum space, and since we are performing an expansion
in the wall velocity, it is natural to use the lowest moments of the
unperturbed particle velocity.}
\beqa
       - v_w \frac{\mu_i'}{T}
           + \ave{\frac{p_z}{\omega }\delta\! f_i'}
        &=&   \aves{C_i}
\nonumber \\
        -\ave{\frac{p^2_z}{\omega^2}} \frac{\mu_i'}{T}
           + v_w \ave{\frac{p_z}{\omega } \delta\! f_i'}
           + v_w \beta \ave{\frac{p_z}{\omega } \delta\! F_i}
        &=&   \ave{\frac{p_z}{\omega }C_i}.
\label{diffeq1}
\eeqa
where the average over momenta is defined as
\beq
\ave{X} \equiv
              \frac{1}{T\,{\int d^{\,3} p \,{\partial 
f_{0}\over\partial\omega}}}
	\left\{ \begin{array}{ll} \int d^{\,3} p \, X \,
	{\partial f_{0}\over\partial\omega}, & X = {p_z^2/\omega^2},\
	(p_z/\omega)\delta\!F_i;\\ 
	\int d^{\,3} p \, X,  & X = {\rm all\ others.} \end{array}
	 \right.
\label{average}
\eeq
In deriving (\ref{diffeq1}) we used the fact that some terms are total
derivatives of momentum, which integrate to zero.
To leading order in the WKB approximation, we
can take $\ave{(p_z/\omega) \delta\! f_i'} \simeq \ave{(p_z/\omega)
\delta\! f_i}'$ since the error is third order in derivatives for
CP-odd quantities, $\sim m'\theta'', m''\theta'$ (fourth order in the
diffusion equation). This lets us express the  l.h.s.\ of (\ref{diffeq1})
in terms of two variables
\beqa
\xi_i &\equiv& \mu_i/T; \nonumber \\
\bvzi &\equiv& \ave{\frac{p_z}{\omega }\delta\! f_i}.
\label{vars}
\eeqa
Writing further $p_z/\omega \equiv v_{p_z}$, (the $z$-component of
the physical group velocity) eq.\ (\ref{diffeq1}) takes a more transparent
form which shows its dependence on powers of the various velocities:
\beqa
\label{diffeq2a}
        -v_w\xi_i' + \bvzi ' &=& \ave{C_i}\\
\label{diffeq2b}
        -\aves{v^2_{p_z}}\xi_i ' + v_w\bvzi ' + v_w \ave{v_{p_z}\delta\! F_i/T}
                          &=& \ave{v_{p_z}C_i}
\eeqa

To proceed, we express the moments of the collision integral appearing in
equations (\ref{diffeq2a}-\ref{diffeq2b}) in terms of the variables $\xi_i$
and $\bvzi$.  This is possible because all perturbations have small
amplitude and we can linearize the collision integrals in $\mu_i$ and
$\delta\! f_i(p,x)$.  In the Appendix it is shown that
\beqa
\label{Cmoment1}
\ave{v_{p_z}C_i} &\simeq&   \bvzi \Gamma^t_i;\\
\label{Cmoment0}
\ave{C_i} &\simeq& - \Gamma^d_{ik} \; \sumxi.
\eeqa
where $\Gamma^t_i$ is the thermally averaged total interaction rate for
the state $i$,  $\Gamma^d_{ik}$ is the averaged interaction rate
corresponding to an inelastic reaction channel labeled by $k$ (we assume
a sum over all allowed interactions) and the abbreviated notation $\sumxi$
in (\ref{Cmoment0}) represents the signed sum over the
chemical potentials of particles participating in that
reaction.\footnote{For example a term induced by the decay process
$i\ra jk$ is more precisely written as $\Gamma_{i\ra jk}\NBR{\xi_i
-\xi_j+\xi_k}$.}

The first moment, eq.\ (\ref{Cmoment1}) is  proportional to the velocity
(kinetic) perturbation. As expected, the relaxation scale for these
perturbations is set by the total interaction rate. The zeroth moment,
on the other hand, only gets contributions from inelastic interactions,
because $\ave{C_{\rm el}}=0$. The damping term appearing in
$\ave{C_i}$ causes relaxation towards chemical equilibrium, which is
reached when the signed sum of the chemical potentials, $\sumxi$,
goes to zero. The relaxation scale is set by the inelastic interaction
rates $\Gamma_{d}$, and is therefore much longer than that of
kinetic perturbations.  These terms also provide the couplings of
particle species which leads to evolution of the asymmetries from the
source species into the left-handed quark fields, a crucial element of
the present baryogenesis mechanism.\\

\noindent {\bf Diffusion equation}.
Differentiating the first moment equation (\ref{diffeq2b}) once, using the
zeroth moment equation (\ref{diffeq2a}) to eliminate $\bvzi$ in the resulting
equation, and neglecting terms of order ${\cal O}(v^2_w)$ (since $v^2_w \ll
\aves{v_{p_z}}$) one obtains the following second order equation for $\xi_i$:
\beq
      - \frac{\aves{v^2_{p_z}}}{\Gamma^t_i}\xi_i''
            - v_w \xi_i'
            - \sum_j \frac{\Gamma^d_{ik}}{\Gamma^t_i} {\xi^{(k)}_j} '
            + \Gamma^d_{ik} \sumxi
            =  -\frac{v_w \beta}{\Gamma^t_i} \ave{v_{p_z}\delta\! F_i}'.
\label{diffeq3}
\eeq
Because of the normalization chosen in (\ref{average}),
(see also (\ref{gammainel})),
an interaction coupling certain species may appear with different effective
strengths in different coupled equations, depending on whether the equation
describes a fermion or a boson chemical potential, and what the particle's
mass is.  For later purposes it will be more
convenient for a given interaction to have the same rate in all
equations.  We therefore multiply each equation by a factor
\beq
\kappa_i \equiv
{\int d^3p\, \partial_{\beta}\left(e^{\beta\sqrt{p^2+m_i^2}}\pm 1\right)^{-1}
     \over	\int d^3p\, \partial_{\beta}(e^{\beta|p|}+1)^{-1} }
\label{newkappa}
\eeq
where $\partial_\beta = {\partial\over\partial\beta}$ and
$\pm$ is  $+$ if particle $i$ is a fermion and $-$ if a boson.
Each interaction rate is then rescaled using
\beq
\tilde \Gamma^d_{ik} \equiv \kappa_i \Gamma^d_{ik}.
\label{redef}
\eeq
The rescaled decay rate is equal  for a given reaction, regardless of
which particle's equation it is appearing in, and the information about
the differences in statistics and masses is encoded in $\kappa_i$'s.
(In the massless limit, $\kappa_i=1$ for fermions and $\kappa_i=2$ for
bosons.) Also, as explained above, $\Gamma^d_{ik}/\Gamma^t_i$ is small
so that we can drop the third term in (\ref{diffeq3}). We then obtain
the usual form of the diffusion equation
\beq
      -\kappa_i (D_i\xi_i '' + v_w\xi ') + \tilde \Gamma^d_{ik}
       \sum_j \xi^\ssk_j = S_i,
\label{diffeq4}
\eeq
where the diffusion coefficient $D_i$, related to the inverse of the
total scattering rate, is given by the usual expression:
\beq
D_i = \frac{\aves{v^2_{p_z}}}{\Gamma^t_i} = \frac{\ave{v^2}}{3}\tau ,
\label{Dcoeff}
\eeq
and the source is
\beq
S_i \equiv - \kappa_i \frac{v_w D_i}{\aves{v^2_{p_z}} T}
                              \aves{v_{p_z}\delta\! F_i'}'.
\label{sources}
\eeq

The proportionality of the force term to the diffusion constant
reflects the fact that the particles must be able to move in response to
the force in order for it to have an effect. The larger the diffusion
constant, the bigger is the response.

\section{Transport in the MSSM}

We now apply the general formalism given above to the MSSM.  The goal
is to see how the asymmetry in source particles due to the chiral force
gets transmitted into left-handed quarks, which drive the sphaleron
interactions to produce a baryon asymmetry.  The network of diffusion
equations connecting the various species of particles in the thermal
bath can be simplified considerably if we take account of the hierarchy
of reaction rates for inelastic processes that change particle
identities.  We take the electroweak sphaleron rate, of order
$\alpha_w^4 T$, to be the slowest of all the relevant interactions,
hence the one we treat last (section 6), whereas the gauge interaction
rates (order $\alpha_i T$) are fast and can be taken to be in
equilibrium on the time scale $D/v_w^2$ for particles to diffuse in
front of the bubble wall.\footnote{
The estimate $D/v_w^2$ can be obtained by setting the distance
a particle diffuses in time $t$ equal to the distance the wall
translates it the same time i.e. $\sqrt{Dt} \sim v_w t$. We will
see that this is indeed the charatceristic scale in the Green's
function (\ref{greenlarger}) we derive below to describe the
diffusion.}

Intermediate between these extremes are certain inelastic processes
that could possibly drive the chiral asymmetry and hence the baryon
asymmetry to zero if they were sufficiently fast. These include strong
sphalerons ($\Gamma_{ss}\sim \alpha_s^4 T$) and helicity-flipping
interactions $\Gamma_{hf} \sim (m^2/T^2) \Gamma_t$, the first of which
tends to erase the chiral asymmetry of the quarks while the second
tends to damp the chiral asymmetry of the charginos which is originally
produced by the classical force.

The only source for the asymmetry in our equations comes from the
chargino sector. It is clear from our results in section 3, that the
wino-like and higgsino-like mass eigenstates have equal and opposite
sources. However, whereas the asymmetry coming from the higgsino-like
states can be efficiently transported to a chiral asymmetry in quarks
and squarks via strong Yukawa interactions, the asymmetry from winos
can only transport via mixing effects through these same interactions.
Indeed, the dominant gauge interactions between winos and quarks do not
give rise to a transport of asymmetry, but produce a large damping term
for the wino asymmetry. Moreover, because the left- and right-handed Winos
have opposite source terms, it is only the difference of their chemical
potentials, $\mu_{\tilde W^+_R} -\mu_{\tilde W^+_L}$, whose diffusion
equation has a nonvanishing source; and the thermal equilibrium of the
supergauge interactions enforces the constraints $\mu_{\tilde W^+_R} -
\mu_{\tilde W^+_L} \propto \mu_{\tilde G} \propto \mu_{\tilde
W^3} \propto \mu_{\tilde B^3}$. Because the neutral gauginos have
helicity-flipping Majorana masses which tend to drive their chemical
potentials to zero, this brings another, gauge-strength suppression
to the wino chemical potential.

For the sake of tractability of the problem, we will here neglect the
chargino mixing effects in the interaction terms, which allows us to
drop the winos from our diffusion equation network. Based on the above
discussion, we believe this to be a very good approximation everywhere
except in the region of parameters where mixing in expected to be
large \ie, when $|m_2| - |\mu| \ll gH_i$.  Even in this case, however,
we err only within the bubble wall where the mixing is large; in front
of the wall where $\ave{H_i} = 0$, there is no mixing.  Since the
diffusion length of the Higgsinos is significantly greater than the
wall width, (see Fig.\ 1(b)) the small mixing approximation is a good
one for most of the region where $\xi_{q_L}$ has support.

Another simplification we have made is to assume that the gaugino
helicity-flip interactions are in equilibrium; this leads to the
constraint that all particle chemical potentials are equal to the
corresponding ones for their superpartners.  Although neither of these
two simplifying assumptions are particularly good approximations, we
expect them to introduce multiplicative errors of at most of order
unity in our estimate of the baryon asymmetry, without changing the
important parametric dependences.

The relevant diffusion equations are most easily written in terms of the
rescaled (s)quark and Higgs(ino) chemical potentials, $\xi_i \equiv \mu_i/T$.
Because of our assumption of vanishing gluino chemical potential and
equilibrium of supergauge interactions, $\xi_{\tilde q}=\xi_q$ for each
light squark species. Moreover, the chemical potentials of quarks within a
doublet are taken to be equal due to weak gauge interactions, for example
$\xi_{b_L} = \xi_{t_L} \equiv \xi_{q_3}$.  The asymmetries between the
particle and antiparticle densities, plus those of their superpartners,
include the $\kappa_i$ factors defined in (\ref{newkappa}):
\beq
\label{Q3eq}
n_{Q_3} - n_{\bar Q_3}
	    =  \frac{T^3}{6}\left(
	        2+\kappa_{\tilde b_L}+\kappa_{\tilde t_L} \right)
\xi_{q_3} \; ,
\eeq
and likewise for the other species.  The factor $2$ in (\ref{Q3eq})
comes from the quarks, while the $\kappa$ factors count the squarks.
(Recall that $\kappa_{\tilde q} = 2$ if the squark $\tilde q$ is light
enough to be in equilibrium in the plasma, and $\kappa_{\tilde q} = 0$
otherwise.) Defining the diffusion operator
\beq
\label{DiffOp}
	{\cal D}_i = - 6\left (D_i \frac{d^2}{dz^2} + v_w \frac{d}{dz} \right)
\eeq
the network of diffusion equations can be written as
\beqa
\label{a1}
        && \hskip 12.5truemm
          {\cal D}_h \xi_{{\tilde h}_1}
	   + \Gamma_{y\mu}(\xi_{{\tilde h}_1}\! -\!\xi_{q_3}\! +\!\xi_{t_R})
	   + \Gamma_{hf}(\xi_{{\tilde h}_1}\! +\!\xi_{{\tilde h}_2})
= S_{H_1} \\
\label{a2}
        &&  \hskip 12.5truemm
           {\cal D}_h \xi_{{\tilde h}_2}
           + (\Gamma_{y}\! +\!\Gamma_{yA})(\xi_{{\tilde h}_2}\!
+\!\xi_{q_3}\! -\!\xi_{t_R})
	   +  \Gamma_{hf}(\xi_{{\tilde h}_2}\! +\!\xi_{{\tilde h}_1})
= S_{H_2} \\
\label{a3}
        && \hskip-13.5truemm
     \sfrac16 (2 \!+ \! \kappa_{\tilde b_L} \!+ \! \kappa_{\tilde t_L}){\cal
D}_q \xi_{q_3}
           + (\Gamma_{y}\! +\!\Gamma_{yA})(\xi_{{\tilde h}_2}\!
+\!\xi_{q_3}\! -\!\xi_{t_R})
           - \Gamma_{y\mu}(\xi_{{\tilde h}_1}\! -\!\xi_{q_3}\! +\!\xi_{t_R})
         \nonumber \\ && \hskip 30.5 truemm
           - \; \Gamma_m \theta(-x) \; (\xi_{t_R} - \xi_{q_3})
           + 2 \Gamma_{ss} = 0 \\
\label{a4}
        &&  \hskip 10.5truemm
           \sfrac12 {\cal D}_q \xi_{t_R}
           - (\Gamma_{y}\! +\!\Gamma_{yA})(\xi_{{\tilde h}_2}\!
+\!\xi_{q_3}\! -\!\xi_{t_R})
           + \Gamma_{y\mu}(\xi_{{\tilde h}_1}\! -\!\xi_{q_3}\! +\!\xi_{t_R})
         \nonumber \\ && \hskip 30.5 truemm
           + \; \Gamma_m \theta(-x) \; (\xi_{t_R} - \xi_{q_3}) -
\Gamma_{ss} = 0 \\
\label{a5}
        && \hskip-14.5truemm
	\sfrac16(2 \!+ \! \kappa_{\tilde d_L}\! +\! \kappa_{\tilde
u_L}){\cal D}_q
	                     \xi_{q_{1,2}}
	   + 2 \Gamma_{ss} = 0 \\
\label{a6}
        && \hskip -4.5truemm
           \sfrac16(1\! +\! \kappa_{\tilde q_R}) {\cal D}_q \xi_{q_R}
           -  \; \Gamma_{ss} = 0,
\eeqa
where $\xi_{q_R}$ stands for any of the right-handed (s)quarks except for
the top.  In section 3.2 we showed that both species of Higgsinos experience
equal chiral forces, so that $S_{H_1} = S_{H_2}$; we will denote it by
$S_H$ henceforth.
The rates $\Gamma_i$ are associated with terms in the interaction Lagrangian
(\ref{Vint}):
\beqa
         \Gamma_{hf} &\leftrightarrow& \mu\tilde h_1 \tilde h_2   \nn\\
	\Gamma_{y} &\leftrightarrow&       h_2 \bar u_R q_L +
          y \bar u_R \tilde h_{2L}\tilde q_L
             + y \tilde u^*_R \tilde h_{2L} q_L  \nn\\
	\Gamma_{y\mu}&\leftrightarrow& y \mu h_1 \tilde q^*_L \tilde
u_R   \nn\\
           \Gamma_{yA} &\leftrightarrow& yA_t\tilde q_L h_2\tilde u^*_R  \
\label{VGamma}
\eeqa
    In addition to these, $\tilde\Gamma_{ss}$ is the rate of strong sphaleron
interactions, which interconvert left- and right-handed quarks of each
flavor; we have used the shorthand
\beq
\label{ssrate}
       \Gamma_{ss} \equiv \tilde\Gamma_{ss} (2 \xi_{q_1} + 2 \xi_{q_2} +
2 \xi_{q_3}
                 - \xi_{u_R}-\xi_{d_R}-\xi_{c_R}-\xi_{s_R}-\xi_{b_R}-\xi_{t_R}),
\eeq
to simplify the appearance of the equations.  Also
$\Gamma_m \theta(-z)$ is the spin-flip rate due to the quark
mass in the broken phase behind the bubble wall, which we take to be
the region $z<0$.  ($\theta(-z)$ is the step function.) There exist
similar gauge-strength interaction terms damping the chemical
potentials $\xi_{{\tilde h}_i}$ in the Higgs condensate region. They could
easily be included in the analysis, but because they are subleading in
strength, we omit them for clarity.

Let us now introduce a ``reference chemical potential,''
$\xi_Q$, which satisfies the equation
\beq
{\cal D}_q \xi_Q - 2\Gamma_{ss} = 0.
\label{refe}
\eeq
Then from (\ref{a5}) and (\ref{a6}), one can express $\xi_{q_{1,2}}$
and $\xi_{q_R}$ in terms of $\xi_Q$:
\beqa
\xi_{q_{1,2}} &=& - \frac{6}{2+\kappa_{\tilde d_L}+\kappa_{\tilde u_L}} \;
           \xi_Q \nn \\
\xi_{q_R}     &=&  \frac{3}{1+\kappa_{\tilde q_R}} \; \xi_Q,
\label{xiQ}
\eeqa
Furthermore, adding (\ref{a3}) and (\ref{a4}) we get the equation
\beq
D_q\left(\xi_{t_R} + \sfrac13(2+\kappa_{\tilde b_L} +\kappa_{\tilde
t_L}) \xi_{q_3}\right)
+ 2\Gamma_{ss}  = 0.
\eeq
This is again of the form (\ref{refe}) and can be used to infer the constraint
\beq
\xi_{t_R} + 2 \tilde \kappa_3 \xi_{q_3}
= - \xi_Q.
\label{xiQdef}
\eeq
where we have defined the frequently appearing quantity
\beq
\tilde \kappa_3 \equiv \sfrac{1}{6}(2 + \kappa_{\tilde b_L} +
\kappa_{\tilde t_L}).
\label{tildekappa}
\eeq
The condition (\ref{xiQdef}) is in fact equivalent to requiring the vanishing
of total baryon number $B_{\rm\sss TOT} \propto \sum_q (1+ \kappa_{\tilde q})
\xi_q \equiv 0$, which is built into our equations until we explicitly
introduce the B-violating electroweak sphaleron interactions.
These constraints leave us with four undetermined potentials, say
$\xi_{t_R}$, $\xi_{q_3}$, $\xi_{{\tilde h}_1}$ and $\xi_{{\tilde h}_2}$.
We can further reduce this number by assuming the strong sphaleron
interactions are in thermal equilibrium.  Using (\ref{xiQ}) and
(\ref{xiQdef}), the strong sphaleron term (\ref{ssrate}) can be expressed
in terms of just $\xi_{t_R}$ and $\xi_{q_3}$:
\beq
\Gamma_{ss} = \tilde \Gamma_{ss} \BBR{
		2(1 +  \tilde \kappa_3 \sigma )\xi_{q_3}
                                         + (\sigma-1) \xi_{t_R} }
\label{ss}
\eeq
where
\beq
\sigma \equiv \sum_{i=1,2}\frac{12}{2+\kappa_{\tilde u_{iL}}+
                                            \kappa_{\tilde d_{iL}}}
                + \sum_{q \neq t} \frac{3}{1+\kappa_{\tilde q_R}}.
\label{eqnS}
\eeq
Assuming strong sphaleron equilibrium  allows us to set (\ref{ss}) to zero,
and thereby determine $\xi_{t_R}$:
\beq
\xi_{t_R}  = 2\frac{(1 +  \tilde \kappa_3 \sigma )}{1-\sigma} \xi_{q_3}
\equiv (1-c_3) \xi_{q_3}.
\label{ssconst}
\eeq

Eq.\ (\ref{ssconst}) must be used with care when eliminating $\xi_{t_R}$
from our system of equations.  If used indiscriminately in (\ref{a3}) or
(\ref{a4}) individually, two incompatible equations would result.  The
correct procedure is to form the linear combination of (\ref{a3}) and
(\ref{a4}) which does not contain the large term $\Gamma_{ss}$.  Eliminating
$\xi_{t_R}$ from this combination gives the correct result at leading order
in an expansion in $1/\Gamma_{ss}$.  In this way we reduce our system of
diffusion equations to a set of just three:
\beqa
    {\cal D}_h \xi_{{\tilde h}_1}
        &+& \Gamma_{y\mu } (\xi_{{\tilde h}_1} - c_3 \xi_{q_3})
         +  \Gamma_{hf} (\xi_{{\tilde h}_1} + \xi_{{\tilde h}_2}) = S_{H}
\label{b1}
\\
    {\cal D}_h \xi_{{\tilde h}_2}
      &+& (\Gamma_y +\Gamma_{yA}) (\xi_{{\tilde h}_2} + c_3 \xi_{q_3})
       +   \Gamma_{hf} (\xi_{{\tilde h}_1} + \xi_{{\tilde h}_2}) = S_{H}
\label{b2}
\\
     \alpha {\cal D}_q \xi_{q_3}
      &-& (\Gamma_y +\Gamma_{yA}) (\xi_{{\tilde h}_2} + c_3 \xi_{q_3})
\nn \\
      &+&  \Gamma_{y\mu } (\xi_{{\tilde h}_1} - c_3 \xi_{q_3})
     - c_3 \, \Gamma_m \xi_{q_3} \; \theta(-z) = 0,
\label{b3}
\eeqa
where $\alpha = 1 -c_3+\tilde \kappa_3$.
Since only the linear combination $\xi_{{\tilde h}_1} + \xi_{{\tilde h}_2}$
is sourced, it is useful to define the linear combinations
\beq
	\xi_{\pm} = \sfrac{1}{2}(\xi_{{\tilde h}_1} \pm \xi_{{\tilde h}_2})
\eeq
Writing equations (\ref{b1}-\ref{b3}) in terms of these variables,
taking the $+$ and $-$ linear combinations of
    (\ref{b1}) and (\ref{b2}), we get
\beqa
{\cal D}_h \xi_+ &+& (2 \Gamma_{hf} + \Gamma_+) \xi_+
                      + \Gamma_- (\xi_- - c_3\xi_{q_3}) = S_H
\label{c1}
\\
{\cal D}_h \xi_- &+& \Gamma_- \xi_+
                      + \Gamma_+ (\xi_- - c_3\xi_{q_3}) = 0
\label{c2}
\\
\alpha{\cal D}_q \xi_{q_3}
                   &+& 2\Gamma_- \xi_+ + 2\Gamma_+ (\xi_- - c_3\xi_{q_3})
                  - c_3 \Gamma_m \xi_{q_3} \theta(-z) = 0.
\label{c3}
\eeqa
where
\beq
\Gamma_\pm \equiv \sfrac{1}{2}\left( \Gamma_{y\mu }
                                 \pm (\Gamma_y + \Gamma_{yA}) \right).
\eeq
These equations show how the chemical potentials for the other combinations
of species, namely $\xi_-$ and the right handed quark asymmetry $\xi_{t_R}$,
are not directly sourced, but develop only after the initial generation of a
nonzero value for $\xi_+$.
Curiously, the quark asymmetry can accidentally vanish if the interaction
rate $\Gamma_-$ is zero, but this is a special case and not generic.
We also see that generation of $\xi_+$ is damped by the fast reaction
rates $\Gamma_+ + 2\Gamma_{hf}$. This is the suppression which has led other
authors to conclude $\xi_+ = 0$ and to therefore ignore the source for this
combination.  The suppression is numerically not very large, however,
because the diffusion length of Higgsinos is comparable to the scattering
length associated with the rates $\Gamma_+ + 2\Gamma_{hf}$.  Thus a
significant asymmetry of ${\tilde h}_1 + {\tilde h}_2$  can build up
despite the damping term in (\ref{c1}).

The $\Gamma_m$ term in eq.\ (\ref{c3}) makes (\ref{c1}-\ref{c3})
difficult to solve. We can proceed by approximating the interaction
terms in eq.\ (\ref{c3}) to be in thermal equilibrium. Then one can
ignore the diffusion operator compared to the damping terms, and solve
for $\xi_{q_3}$
\beq
\xi_{q_3} = \frac{1}{c_3}
            \NBR{\frac{\theta(-z)}{(\mhalf \Gamma_m + \Gamma_+)}
                         +  \frac{\theta(z)}{\Gamma_+}}
                    \nBR{\Gamma_-\xi_+ + \Gamma_+ \xi_-}.
\label{xitr}
\eeq
(In fact only the region $z>0$ will be relevant for what follows.)
The approximation can be justified by estimating the relative sizes
of ${\cal D}_q\xi_{q_3}$ and the rate terms.  Below we will show that
$\xi_{q_3} \sim \exp (-k_h x)$, where $k_h^2 \sim
\Gamma_+/6D_h$ for large $\Gamma_+$. Thus
\beq
{\cal D}_q \xi_{q_3} \sim D_q k_h^2 \xi_{q_3}
                           \sim \Gamma_+ \frac{D_q}{D_h} \xi_{q_3}
                           \ll \Gamma_+ \xi_{q_3},
\label{estimate}
\eeq
A subtle point here is the use of
the large diffusion length $k_h^{-1}$ of the Higgsinos, rather than
the much shorter one which might be expected for quarks, due to their
stronger interactions.  Physically this occurs because the diffusion tail
of $\xi_{q_3}$ comes from the tranformation of efficiently-diffusing
Higgsinos into quarks, by interactions like ${\tilde h} + {\tilde t}_R
\to q_3$.

Using (\ref{xitr}) in eqs.\ (\ref{c1}-\ref{c2}), the problem is reduced
to two coupled equations:
\beqa
{\cal D}_h \xi_+ &+& \NBR{ 2\Gamma_{hf} + \Gamma_+ - \Gamma_-^2
                        \NBR{\frac{\theta(-z)}{\Gamma_+ + \mhalf \Gamma_m}
                           + \frac{\theta(z)}{\Gamma_+} } }\, \xi_+
         \nn\\  &+& \theta (-z) \frac{\Gamma_- \Gamma_m}{2 \Gamma_+
                + \Gamma_m}\xi_- = S_H
\label{s2a}
\\
{\cal D}_h \xi_- &+& \frac{\Gamma_+ \Gamma_m}{2 \Gamma_+ + \Gamma_m}
                       \,\theta (-z) \xi_-
                  = - \frac{\Gamma_- \Gamma_m}{2 \Gamma_+ + \Gamma_m}
                       \,\theta (-z)\, \xi_+.
\label{s2b}
\eeqa
We note that $\xi_-$ is generated only in the broken phase, behind the
bubble wall.  If the top quark helicity flip rate $\Gamma_m$ was zero,
$\xi_-$ would decouple completely and the system would be reduced
to a single equation for $\xi_+$, easily solvable by Greens
function methods. However the helicity flip rate is large, since
$m_t\sim y v_c/T_c \sim 1$, where $v_c$ is the Higgs VEV at the critical
temperature $T_c$.  Thus (\ref{s2a}-\ref{s2b}) cannot be further
simplified without additional assumptions.  One such assumption which
is quite plausible is that the ratio
\beq
           R \equiv \Gamma_-/\Gamma_+
\eeq
is small, in which case $\xi_+$ and
$\xi_-$ do approximately decouple.  Even  with a
rather mild suppression $R \lsim 1/2$, we can drop the $\xi_-$
and $\Gamma_-$-terms in eq.\ (\ref{s2a}).  The solution then proceeds in
two steps.  First one solves eq.\ (\ref{s2a}) for $\xi_+$,
which now reads
\beq
{\cal D}_h \xi_+  + (2\Gamma_{hf} + \Gamma_+)  \xi_+ = S_H.
\label{f1}
\eeq
This can be done using the Greens function:
\beq
\xi_+(z) = \frac{1}{6} \int_{-\infty}^\infty dy\, G(z-y) S_H(y),
\label{xih}
\eeq
where
\beqa
\label{Green}
            G(x) &=& \frac{D_h^{-1}}{k_+-k_-}
              \left\{\begin{array}{cl}
                  e^{-k_+x}, & x>0  \\
                  e^{-k_-x}, & x<0
                      \end{array}
                   \right. \nonumber\\
           k_\pm &=& \frac{v_w}{2D_h}
                \left(1\pm \sqrt{1 + \frac{2 \bar \Gamma D_h}{3v_w^2}} \right);
           \qquad  \bar \Gamma \equiv 2\Gamma_{hf} + \Gamma_+.
\label{Greeni}
\eeqa
Although $\xi_-$ is suppressed relative to
$\xi_+$ by the factor $R$,  one cannot simply neglect it, because in
eq.\ (\ref{xitr}) for $\xi_{q_3}$ the contribution from
$\xi_-$ is enhanced by $1/R$ relative to that from $\xi_+$.  In our
approximation the equation (\ref{s2b}) for $\xi_-$ becomes
\beq
{\cal D}_h \xi_-  + \frac{\Gamma_+ \Gamma_m}{2 \Gamma_+ + \Gamma_m}
                       \,\theta (-z)\, \xi_-
                 =  - \frac{\Gamma_- \Gamma_m}{2 \Gamma_+ + \Gamma_m}
                       \,\theta (-x)\,  \xi_+,
\label{xismH}
\eeq
where $\xi_+$ is the integral solution (\ref{xih}).
Defining a scaling parameter
\beq
\gamma \equiv 2 + \frac{\Gamma_m}{\Gamma_+}
\label{scaling}
\eeq
equation (\ref{xismH}) may be written as
\beq
\gamma {\cal D}_h \xi_- + \Gamma_m \theta(-x) \xi_-
        = - R \, \Gamma_m \theta(-z) \xi_+,
\label{xismH2}
\eeq
which can also be solved by Greens function methods. We
only need to know the solution in the symmetric phase, because
the sphalerons which are being biased by $\xi_{q_3}$ are highly
suppressed in the broken phase. We find
\beq
\xi_-(z) = - \sfrac{1}{6} R\, \Gamma_m \int_{-\infty}^0
                     dy\, G_>(z,y) \xi_+(y)\; ,
        \qquad  z > 0,
\label{xihres}
\eeq
where the new Greens function is
\beqa
G_>(z,y) &=& \frac{1}{\gamma v_w}\; e^{- v_w z/D_h}
                \left\{  - \theta(-y) \frac{v_w}{\alpha_- D_h}
                           e^{- \alpha_-y} \right.
                           \nn \\
          &+&       \left. \theta(y)\theta(z-y)
               \left(e^{v_w y/D_h} + \frac{\alpha_+}{\alpha_-} \right) \right.
                \left. + \theta (y-z) \left(e^{v_w z/D_h} +
                        \frac{\alpha_+}{\alpha_-} \right) \right\}
\label{greenlarger}
\eeqa
(of which only the first term enters in (\ref{xihres})) with
\beq
\alpha_\pm = - \frac{v_w}{2D_h}\left( 1 \mp \sqrt{1 +
                    \frac{2\Gamma_m D_h}{3 \gamma v_w^2}} \right).
\label{alphapm}
\eeq
It can be seen that the solution $\xi_-$ vanishes not only when
$R \ra 0$, but also when $\Gamma_m \ra 0$, as it should. Inserting the
solution (\ref{xih}) for $\xi_+(y)$ into (\ref{xihres}) and performing
the $y$ integral, one can write $\xi_-$ as
\beq
\xi_-(z) = - \sfrac{1}{36} R \frac{\Gamma_m}{\alpha_- \gamma D_h}
      e^{-(v_w/D_h + k_B)z} \, \int_{-\infty}^\infty dy\,  {\cal G}_-(y)
	\, S_H(y),
\qquad  z>0,
\label{xihfinal}
\eeq
with the kernel ${\cal G}_-$
\beqa
{\cal G}_-(y) &=& \frac{D_h^{-1}}{k_+-k_-}
           \left\{ \theta (y) \frac{1}{\alpha_-+k_-} e^{k_-y} \right.
           \nn \\ && \hskip 0.8truecm \left.
           +\theta (-y) \left( \frac{1}{\alpha_-+k_-} e^{-\alpha_-y}
                            + \frac{1}{\alpha_-+k_+} (e^{k_+y}
                                                    - e^{-\alpha_-y})
                      \right) \right\}.
\label{calG1}
\eeqa

With these solutions for the Higgsino chemical potentials, eq.\ (\ref{xitr})
gives that of the third generation left-handed quarks.  The first and
second generation quark potentials are determined by eqs.\
(\ref{xiQ}, \ref{xiQdef}, \ref{ssconst}).  We are now ready
to consider how the quarks bias sphalerons to produce the baryon
asymmetry.

\vskip 0.5truecm

\subsection{Baryon asymmetry}

Local baryon production is sourced by the total left-handed quark and
lepton asymmetries in front of the bubble wall. In the present scenario,
there is essentially no lepton asymmetry.
Thus the source for baryon production by the passing wall is just the
left-handed quark asymmetry,
$\xi_{q_L}$, which enters the baryon violation rate equation as
\beq
     \frac{\partial n_B}{\partial t}
              = \sfrac32 \Gamma_{\rm sph} \left( \xi_{q_L}
              -  A \frac{n_B}{T^2} \right).
\label{Beqn}
\eeq
Here $\Gamma_{\rm sph} \equiv \kappa_{\rm sph} \alpha_W^5 T^4$ is the
Chern-Simons number (CSN) diffusion rate across the energy barrier which
separates $N$-vacua of the SU(2) gauge theory, where $\kappa_{\rm sph} =
20\pm 2$ \cite{MooreSp}.
The second term describes sphaleron-induced relaxation of the baryon
asymmetry in the symmetric phase (more about which below).
Using (\ref{xiQ}, \ref{xiQdef}, \ref{ssconst} and \ref{xitr}) one finds
that the quark chemical potential created by the classical CP violating
force in the wall, combined with fast Yukawa and strong sphaleron processes,
is
\beqa
      \xi_{q_L}
          &=&   3 ( \xi_{q_1} + \xi_{q_2} +  \xi_{q_3} )
\nn \\
          &=&   3\, C(\tilde \kappa_i) \,
               \nBR{R \xi_+ + \xi_-},
      \qquad z > 0.
\label{beeL}
\eeqa
where the factor of 3 counts the quark colors,
$\xi_+ (x)$  and $\xi_-(x)$ are given by eqs.\
(\ref{xih}) and (\ref{xismH}), and
\beq
C(\tilde \kappa_i) \equiv
              \frac{1}{c_3}\NBR{\NBR{\frac{1}{\tilde \kappa_{1}}
                                   + \frac{1}{\tilde \kappa_{2}}}
                            \nBR{1+2 \tilde \kappa_3  - c_3} + 1},
\label{ccoeff}
\eeq
with $\tilde \kappa_{1,2}$ defined analogously to $\tilde \kappa_3$ in
eq.\ (\ref{tildekappa}), and $c_3$ given by (\ref{ssconst}).  The
coefficient $C(\tilde \kappa_i)$ encodes the essential information
about the effect of the squark spectrum on our results, as will be
shown below. We remind the reader that (\ref{beeL}) is valid to leading
order in an expansion in $R = \Gamma_-/\Gamma_+$, which is assumed to
be smaller than unity; recall that $\xi_-$ is of order $R$ in
(\ref{xihfinal}).

The second term on the r.h.s.\ of (\ref{Beqn}) is the Boltzmann term
which would lead to relaxation of the baryon number if the sphaleron
processes had time to equilibrate in front of the bubble wall.  This
would be the case if the bubble wall was moving very slowly.
Thus the $n_B$ appearing here is related to the left-handed quark
and lepton asymmetries, $\mu_q$ and $\mu_l$,
that would result from equilibrating all
flavor-changing interactions which are faster than the sphaleron rate
in the symmetric phase.   Thus $A$ is given by
\beq
A \frac{n_B}{T^2} \equiv \mu_{CS} = 9 \mu_q + \sum \mu_{l_i},
\label{Adef}
\eeq
where $\mu_{CS}$ is the chemical potential for Chern-Simons number,
and the latter equality
follows from the fact that each sphaleron creates nine quarks and three
leptons.  Because of efficient quark mixing, all
quarks have the same chemical potential $\mu_q$. In the leptonic sector
however, the mixing may be weak (depending on the neutrino and slepton
mass matrices) and each flavor asymmetry may be separately conserved.
To solve for these chemical
potentials, one must determine which interactions are in equilibrium on the
relevant equilibraton time scale, which depends on the spectrum
of supersymmetric particles carrying baryon and lepton number.  In the
usual wash-out computation in the broken phase, the appropriate time scale
is the inverse Hubble rate, which means that even the feeblest Yukawa
interactions leading to $e_R$-equilibration \cite{CKO} are considered
to be fast. In this case, using the notation
\beq
           n_a\equiv  N_a - N_{\bar a} = \kappa_a \frac{\mu_a T^2}{6},
\eeq
where $\kappa_a =1 (2)$ when $a$ refers to a fermion (boson), one can
show that for the SM
\beq
n_B = \sfrac{1}{3} n_q = \frac{1}{3}(6 \times 3 \times 2)\frac{\mu_q T^2}{6}
      \; \Rightarrow \; \mu_q = \sfrac12\, {n_B\over T^2}
\eeq
and similarly $\sum_i \mu_{l_i} = 2n_B/T^2$, which, when inserted in
(\ref{Adef}) gives the familiar result $A = 13/2$.
For electroweak baryogenesis, however, the relevant time scale is the
inverse sphaleron rate in the symmetric phase, and therefore none of the
right-handed leptons will have time to equilibrate ($\tau$ is in fact a
border-line case with chirality flipping rate comparable to the sphaleron
rate; but we take it also to be out of equilibrium).
Then, with the SM spectrum, which would apply if all squarks were heavy,
$\sum_i \mu_{l_i} = 3n_B/T^2$ and hence $A = 15/2$.  If there are
$N_{sq}$ flavours squarks which are light enough to be present at $T=100$
GeV, one has
\beq
A =  \frac{9}{2}\, \left(1 + \frac{N_{sq}}{6}\right)^{-1} + 3.
\label{finMSSMbroken}
\eeq
It is straightforward to generalize $A$ to the case of an arbitrary number of
light left-handed sleptons, but the expression is cumbersome because
of the multitude of possible mixing scenarios in the leptonic sector, and we
omit it here for the sake of simplicity.

Moving to the wall frame, the time derivative in (\ref{Beqn}) becomes
$\partial_t \ra-v_w\partial_z$, and it is easy to
integrate the equation to obtain the baryon asymmetry:
\beq
n_B =  \frac{3\,\Gamma_{\rm sph}}{2\,v_w} \int_0^\infty dz\,
            \xi_{q_L}(z) e^{-k_B z},
\label{Btotal}
\eeq
where
\beq
     k_B \equiv \frac{3 A}{2 v_w}\, \frac{\Gamma_{\rm sph}}{T^3}.
\label{keibii}
\eeq
\vskip 0.5cm
The integral over $z$ in (\ref{Btotal}) can be done
analytically. The baryon-to-entropy ratio, $\eta_{\sss B} \equiv n_B/n_\gamma
\simeq 7n_B/s$, can then be written as a single integral over the source
function $S_H(y)$:
\beq
     \eta_{\sss B} =
         \frac{945 \kappa_{\rm sph} \alpha_W^5}{8\pi^2 v_w g_*} \;
          C(\tilde \kappa_i) \; R \; \int_{-\infty}^\infty
           dy\,  \NBR{{\cal G}_+(y)
                 - \frac{\Gamma_m}{6 \gamma \alpha_- (v_w + D_hk_B)}
                                            {\cal G}_-(y)}\, S_H(y),
\label{bnumber}
\eeq
where we have scaled the variable $y$ to units $1/T$, and the new kernel,
arising from performing the $z$-integral over $\xi_+(z)$, is given by
\beq
{\cal G}_+(y) = \frac{D_h^{-1}}{k_+-k_-}
                 \left\{ \theta(y)\NBR{\frac{e^{k_-y}}{k_- + k_B}
                          +  e^{-k_By}\NBR{\frac{1}{k_+ + k_B}
                                          -\frac{1}{k_- + k_B}}}
                     +\theta(-y) \frac{e^{k_+y}}{k_+ + k_B} \right\}.
\label{calG2}
\eeq
The ratio of the contributions coming directly from $\xi_+$ and from
the indirectly sourced $\xi_-$ is controlled by the parameter
$\Gamma_m/(\alpha_-(v_w + D_h k_B))$.
Using typical values for the other parameters,
the $\xi_-$ term turns out to be significant for $v_w \simeq 0.01$,
and subdominant for larger or smaller wall velocities.

\subsection{Sources}

We still need to calculate the source $S_H$ appearing in
the above equations.  It is given by the thermal average (\ref{sources})
where the force $\delta\! F$ corresponds to the CP-violating part of the
classical force, eq.\ (\ref{chforce}).  For a Higgsino of helicity
$\lambda$, using $\kappa_{\tilde h} = 1$ since Higgsinos are fermions,
we have
\beq
S_H = -	\frac{\lambda}{2}\frac{v_w D_h}{\aves{v_{p_z}^2}T} \,
             \ave{\frac{|p_z|}{\omega^3}} \, (m_\pm^2 \theta_\pm')'',
\label{finalsc}
\eeq
where the sign $\pm$ is defined to be the sign of $|\mu|-|m_2|$,
since the lighter (heavier) of the local mass eigenstates $m^2_{\pm}$ is
the Higgsino-like particle when $\mu < m_2$ ($\mu > m_2$).  The absolute
value on $|p_z|$ comes from the relation between spin and helicity:
$s p_z = \lambda |p_z|$. The average $\aves{v_{p_z}^2}$ is very
accurately approximated by the fit
\beq
      \aves{v_{p_z}^2} \cong \frac{3x_\pm + 2}{x_\pm^2 + 3x_\pm + 2},
\eeq
where $x_{\pm} \equiv m_\pm/T$, and using Maxwell-Boltzmann statistics,
one can show that
\beq
\ave{\frac{|p_z|}{\omega^3}} =
            \frac{(1 - x_\pm)e^{-x_\pm} + x_\pm^2E_1(x_\pm )}
                 {4 m_\pm^2 K_2(x_\pm)},
\eeq
where $K_2(x)$ is the modified Bessel function of the second kind and
$E_1(x)$ is the error function \cite{abste}.

In deriving (\ref{finalsc}), we implicitly assumed that the
charginos are
light compared to the temperature. In the limit that they
become heavy they must decouple however, and as a result the damping
rates $\Gamma_\pm$ for Higgsinos to be transformed into quarks/squarks,
in eq.\ (\ref{c3}), must go to zero.
The approximations (\ref{xitr}) and (\ref{estimate}) would consequently
break down. In an exact treatment, one should solve the equations
(\ref{c1}-\ref{c3}) numerically in these cases.  We will instead adopt a
simpler approximation, incorporating the effect of decoupling
on the chargino source $S_H$ with a suppression factor
$n(m_{\tilde h})/n(0)$, that is, the ratio of thermal densities for
a particle of mass $m_{\tilde h}$ relative to a massless particle.
In this approximation, $S_H$ becomes
\beq
S_{H,\rm eff} = -\frac{s}{4}\frac{v_w D_h}{{\aves{v_{p_z}^2}T^3}}
                   \NBR{ (1 - x_\pm) e^{-x_\pm} + x_\pm^2E_1(x_\pm) }
                   \; (m_\pm^2 \theta_\pm')''.
\label{effsc}
\eeq
The effect of this modification is small for chargino masses up to
200 GeV. Beyond this it becomes crucial for suppressing
baryon production
from particles too heavy to be present in the thermal bath.
For $|\mu | \gsim 200$ GeV, our results should be understood to
have a multiplicative uncertainty of order unity arising from this
approximation.

\begin{figure}[t]
\centering
\vspace*{-12mm}
\hspace*{-6mm}
\leavevmode\epsfysize=10.5cm \epsfbox{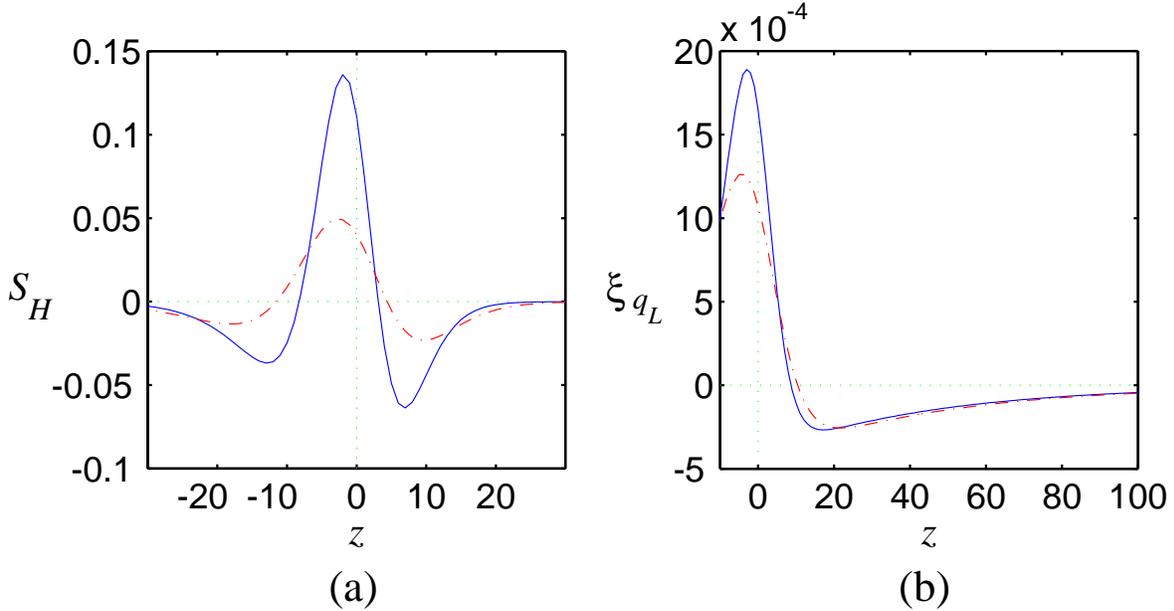}\\[-6mm]
\caption[fig1]{\label{fig1} (a) The source for baryogenesis from the
chiral classical force, eq.\ (\ref{effsc}), for the parameters $\mu =
100$ GeV and $m_2 = 150$ GeV and $\ell_w = 10/T$ (solid line) and
$\ell_w = 14/T$ (dashed line). (b) The left-handed quark asymmetry
$\xi_{q_L}$, eq.\ (\ref{beeL}), for the same parameters. The distance
from the center of the wall $z$, is measured in units $1/T$.}
\end{figure}

To fully specify the source term, we must also give the functional form
for the spatial variation of the Higgs field condensate, since this
enters the  mass eigenstates $m_\pm$ and the CP-violating phase
$\theta_\pm^\prime$ in the above formulas.  Because our source is
proportional to the combination $H_1 H_2'+ H_2 H_1'$ of the two Higgs
fields, our results are not sensitive to changes in the ratio
$\tan\beta = H_2/H_1$, which is in fact known to be nearly constant for
bubble walls in the MSSM---at least for generic parameter values,
including those that give a strong enough phase transition.  This is
in marked contrast to other analyses where the source was assumed to
be proportional to $H_1 H_2'-H_2 H_1'$.  It thus suffices for us to use a
simple kink profile
\beq
u(z) \equiv g\sqrt{H_1^2 + H_2^2} = g\,\frac{v_c}{\sqrt{2}}\,
	\frac{1}{2}\left(1 - {\tanh}\left(\frac{z}{\ell_w}
	\right)\right)
\eeq
with $u_1 = u\sin \beta $ and $u_2 = u\cos \beta$. Here $\ell_w \sim
6-14/T$ \cite{MQS,CM} is the wall width and $v_c$ is the value of the
Higgs condensate at the critical temperature. This VEV has the usual
normalization in the vacuum $v_{T=0} \simeq 246$ GeV, while the
requirement for a strongly enough first order transition (to avoid
washout) is $v_c/T_c > 1.1$.

We display the source (\ref{effsc}) as a function of position relative
to the wall (at $z=0$) in Fig.\ 1(a), using the parameters values $m_2
= 150$ GeV $\mu = 100$ GeV, $\delta_\mu \equiv \arg(m_2\mu ) = \pi/2$
and  $v_w = 0.3$ for two different wall widths: $\ell_w = 10/T$ and
$\ell_w = 14/T$. For the parameters left unspecified above we use the
following standard reference values:
\beqa
&& D_h = 20/T \quad \Gamma_{hf} = 0.013 \quad \Gamma_m = 0.007 T
      \quad \Gamma_+ = 0.02 T, \quad R = {\Gamma_-\over\Gamma_+}= 0.25 \nn \\
&&  \ell_w = 10/T,  \quad v_c = 120,  \quad T=90,
      \quad \tan \beta = 3.
\label{stdset}
\eeqa

In the limit that $u^2 \ll \mu^2, m_2^2$, the source would be a symmetric
function of $z$ since it would be proportional to $(u^2)'''$.  However for
finite $\mu$ and $m_2$, the actual $z$-dependent mass eigenvalues appearing
in the coefficient of $(u^2)'''$ depend on $u^2$ rather than its derivatives,
hence the departure from the symmetric form. In figure 1(b) we plot the
profile for left-handed quark number, $\xi_{q_L}$, for the same set of
parameters.  As expected, the spatial extent of the quark asymmetry is
roughly the diffusion length of the Higgsinos,  $D_h/v_w\sim 60/T$ (see
the remarks below eq.\ (\ref{estimate})). For a smaller wall velocity the
diffusion tail extends further, but the amplitude of $\xi_{q_L}$ in the
tail gets smaller because then the damping has more time to suppress the
chargino asymmetry.\\

The rates quoted in (\ref{stdset}) are rough estimates, obtained from an
approximate computation of a subset of relevant $2\ra 2$ reaction rates,
and higgsino decay rates, when kinematically allowed. For example $\tilde
h_2 W^3_\mu \ra \tilde t_R q_L$, gives
\beq
\sigma_W \simeq \frac{g^2y^2}{64\pi s}\nBR{\frac{3}{2}
+ {\rm ln}\frac{s}{m_{\tilde t_R}^2}},
\eeq
where $s \simeq 20T^2$ is the center of mass energy squared and $m_{\tilde t_R}^2 \simeq
m_U^2 + 0.9T^2$ is the thermal mass of the right handed squark \cite{ComEs}.
The soft SUSY breaking mass parameter $m_U^2$ is taken to be negative, $m^2_U
\sim -60^2$ GeV$^2$, as indicated by the need to get a strong enough first
order phase transition \cite{LR}. The rates 
(\ref{stdset}) correspond to a conservative overestimate by a factor of
5 over the total averaged contribution from various scattering
channels (for how to perform the thermal averages, see ref.\ \cite{CKO}).
The decay rates have a fairly strong dependence on higgsino mass $m_{\tilde h}
\simeq \mu$. For example $\tilde h_{2_L} \ra \tilde t_R b_L^c$ gives
\beq
  \Gamma \simeq \frac{m_{\tilde h}}{16\pi }
  \lambda^{3/2}(1,\sfrac{m_{b_L}^2}{m_{\tilde h}^2},
                   \sfrac{m_{\tilde t_R}^2}{m_{\tilde h}^2}),
\eeq
where $\lambda (x,y,z) \equiv (x - y - z)^2 - 4yz$ and $m_{b_L}\simeq 0.76T$
\cite{DKO}. For small $\mu \lsim 130$ GeV, the decay channels are not open,
whereas for large enough $\mu$ they dominate over the scattering contribution.
However, our numerical results for $\eta_B$ are fairly insensitive to changes
in various rates; for example decreasing $\Gamma_+$ by a factor of 5,
appropriate for $\mu \lsim 130$ GeV would increase $\eta_B$ by about 30 per
cent, whereas incrasing it by a factor of 5, appropriate for $\mu \simeq 500$
GeV, would decrease $\eta_B$ by about 40 per cent.  This scaling is somewhat
weaker than the naively expected $\eta_B \sim 1/\sqrt{\Gamma_+}$ dependence,
because the damping effect due to faster rates is initially being
compensated by more efficient transport from the chargino sector. (We have
checked that the naive scaling eventually follows for values of $\Gamma_+$
large enough that the transport effect has been saturated.)  Nevertheless, the
relative insensitivity of the results on the rates warrants our use of
the rough estimates (\ref{stdset}) in our numerical work.
\\

\section{Results}

\noindent{\bf Dependence on squark spectrum.} Let us first consider the
dependence of $\eta_{\sss B}$ on the squark spectrum. This is contained
in the parameter $C(\tilde \kappa_i)$, some representative values of
which are given in Table 1.  For certain choices of squark masses,
$C(\tilde \kappa_i)=0$, which reflects the approximation we made of
taking the strong sphalerons to be in equilibrium; it is well known
that these interactions tend to damp the baryon asymmetry if, for
example, no squarks are present \cite{GS}. In these cases the baryon
asymmetry is not really zero, but comes from
$1/\Gamma_{ss}$ corrections which we have not computed.  Ignoring such
corrections, one sees the clear preference for the minimal possible
number of light squark species from $C(\tilde \kappa_i)$.  This is
fortuitous because it coincides with the need for a single, light,
right-handed stop in order to get a strong phase transition.  If the
left-handed stops and sbottoms are also light (which, incidentally, is
incompatible with the large radiative corrections needed for the Higgs
mass to satisfy the experimental lower limit, as well as rho parameter
constraints) the baryon asymmetry is reduced by a factor of ten.  Thus,
considerations both of the initial baryon production and the
preservation from washout favor the ``light stop scenario.''\ \ Since
the effects of the spectrum are trivial to account for in the final
results, being just an overall multiplicative factor, we shall
henceforth concentrate only on the most favorable scenario. \\

\begin{table}[t]
\begin{center}

\begin{tabular}{|l|c|} \hline
{\em light squarks} & $C(\tilde \kappa_i)$
\\
\hline\hline
\hskip 1truemm
All             &   0    \\ \hline
All R-chiral    &   0    \\ \hline
All 3rd family  &   0    \\ \hline
$\tilde t_L, \; \tilde b_L$ and $\tilde t_R^{\phantom{|}}$
                    &  2/41 \\ \hline
$\tilde t_L$ and $\tilde t_R^{\phantom{|}}$
                    &  3/16 \\ \hline
$\tilde t_R$ and $\tilde b_R^{\phantom{|}}$
                    &  3/8 \\ \hline
$\tilde t_R^{\phantom{|}}$ only
                    & 10/23
\\ \hline
\end{tabular}

\vskip 2mm
\caption{Multiplicative factor $C(\tilde \kappa_i)$ containing the
dependence of $\eta_{\sss B}$ on the squark masses for particular choices of
the light squark spectrum.}

\end{center}
\end{table}

\noindent{\bf Velocity dependence.}  The dynamics of the phase
transition, even apart from CP-violating effects studied here, is a very
complicated phenomenon, involving hydrodynamics of the fluid
interacting with the expanding walls, and reheating effects due to the
latent heat released in the transition \cite{bubexp}.  Although the
originally spherical bubbles quickly grow and reach some terminal
velocity, inhomogeneities can subsequently develop. This occurs when
the shock waves from the bubble expansion heat the ambient plasma and
thereby reduce the latent heat released as regions of space are
converted from the symmetric to the broken phase.  There is a
subsequent decrease of pressure driving the expansion, and depending on
model parameters, may lead to significant slowing down of the walls.
The process of heating by a collection of shock waves causes local
variations in the temperature as well as fluid velocities, with
consequent deformation of the shape and speed of the wall. These
variations occur on the macroscopic length scale of the bubble radius,
which is many orders of magnitude greater than the microphysical scales
that have been discussed here so far. In this sense, eq.\ (\ref{Btotal})
gives only the {\em local} baryon number at a given position in space
after the wall passes by. The presently observed asymmetry should be
computed by averaging over a region which is large compared to the
bubble size at the time the phase transition completes:
\beq
\eta_{\sss B} = \frac{1}{{\rm Volume}}\int d^{\,3}{x}\,
                   \eta_{\sss B}[v_w(x)],
\label{bave}
\eeq
where $\eta_{\sss B}$ is considered as a functional of the locally varying
wall velocity.
Only if the phase transition is very strong, so that there is a high
degree of supercooling, will the reheating effects leading to
inhomogeneities be small or negligible.

\begin{figure}[t]
\centering
\vspace*{-12mm}
\hspace*{-4mm}
\leavevmode\epsfysize=10.5cm \epsfbox{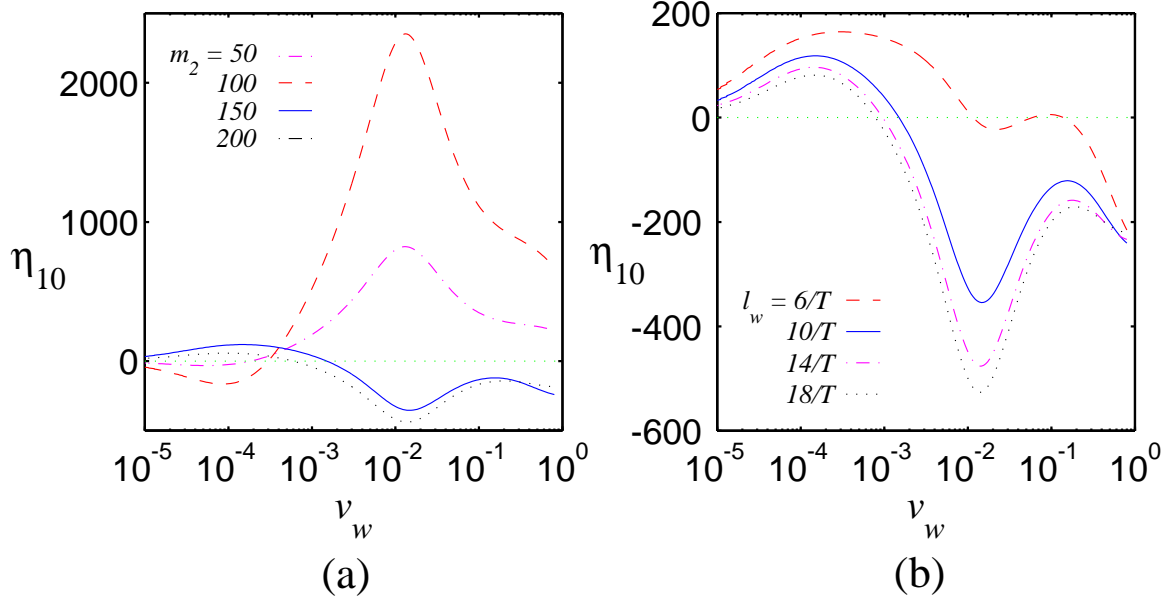}\\[-3mm]
\caption[fig2]{\label{fig2} $\eta_{10}$ as a function of wall velocity
for $\mu = 100$ GeV and $\sin \delta_\mu=1$ for: (a) a varying gaugino
mass parameter $m_2 =50$, $100$, $150$ and $200$ GeV and: (b) for
a varying wall width, $\ell_w = 6/T$, $10/T$, $14/T$ and $18/T$.}
\end{figure}

In addition to the possibility that $v_w$ has spatial inhomogeneities,
it is also interesting to study the dependence on $v_w$ simply because
its value is not yet known with great certainty, although some progress
has recently been made \cite{Moore,JS}.  Our treatment takes into
account the back-reaction effect on the baryoproduction (washout by
sphalerons), so our results are valid for arbitrarily small wall
velocities. In Fig.\ 2(a) we plot $\eta_{10} \equiv \eta_{\sss B}
\times 10^{-10}$ as a function of $v_w$ for $\mu = 100$ GeV and
$m_2 =50$, $100$, $150$ and $200$ GeV, and in Fig.\ 2(b) for
four different values of the wall width $\ell_w$, with $\mu=100$
GeV and $m_2=150$ GeV.

The peak occurring at $v_w \simeq 0.01$ for some parameters in Fig.\ 2
(a), first observed in \cite{CK}, is due to the contribution from the
${\cal G}_-$ term in (\ref{bnumber}). This is enhanced by a factor
$(v_w + D_h k_B)^{-1}$, which for the assumed parameter values peaks
near $v_w \lsim \sqrt{D_hk_B} \simeq 0.01$.  Because of the
back-reaction, the baryon asymmetry vanishes when the wall velocity
goes to zero.  The peak is prominent only for the values of $m_2 \sim
\mu$ however, and the typical velocity depencence of $\eta_{10}$ is not
quantitatively very large as a function of velocity. It is quite
complicated however, in that for special parameter values the
asymmetry  can accidentally be small or zero. The crossings through
zero arise as follows: for relatively large $v_w$ the baryon production
in the diffusion tail dominates over the opposing contribution
generated near the wall (see the generic form of the $\xi_{q_L}$
distribution in Fig.\ 1 (b)). For small wall velocities the length of
the diffusion tail increases as $D/v_w$, but the amplitude of the
asymmetry gets smaller due to interactions, which have more time to
damp the asymmetry. Moreover, the contribution from the part of the
diffusion tail extending beyond $1/k_B$ is cut out, because the baryon
asymmetry is already relaxing due to sphaleron washout beyond that
distance.  As a result the contribution from the tail eventually
becomes the smaller one, leading to a cancellation between the two
contributions that give the net asymmetry.

While the uncertainty in $v_w$ at present is not necessarily the dominant
one for estimating the baryon asymmetry, determining $\eta_{10}$ to high
precision for a given set of chargino mass parameters would need careful
hydrodynamical modelling of the bubble wall expansion. Also, even rather
small fluctuations in $\eta_B$ can have interesting consequences
elsewhere: for example they can seed the generation of large fluctuations
in leptonic asymmetries in certain neutrino-oscillation models \cite{EKS}
with potentially large effects on nucleosynthesis.\\

\noindent{\bf Dependence on chargino mass parameters.} The most
important supersymmetric inputs directly affecting the  baryon
asymmetry are the chargino mass parameters $m_2$ and $\mu$, and the
CP-violating phase $\delta_\mu\equiv {\rm arg}(m_2\mu)$. In Fig.\ 4a
and 4b we plot the contours of constant $|\delta_\mu|$  giving the
desired baryon asymmetry $\eta_{\sss B} = 3 \times 10^{-10}$ \cite{FKOT}
in the $(m_2,\mu )$ plane. The baryoproduction is most efficient for small
masses, $m_2, \mu \lsim 100$ GeV, but baryons can still be copiously
produced for $|m_2|$ and $|\mu |\sim$ 500 GeV and higher. In the best
cases, a large enough baryon asymmetry can be produced even with a
very small explicit CP-violating angle of order a few $\times 10^{-3}$,
comfortably within the constraints coming from electric dipole moment
searches \cite{dmom}.

\begin{figure}[t]
\centering
\vspace*{-4mm}
\hspace*{-6mm}
\leavevmode\epsfysize=10.5cm \epsfbox{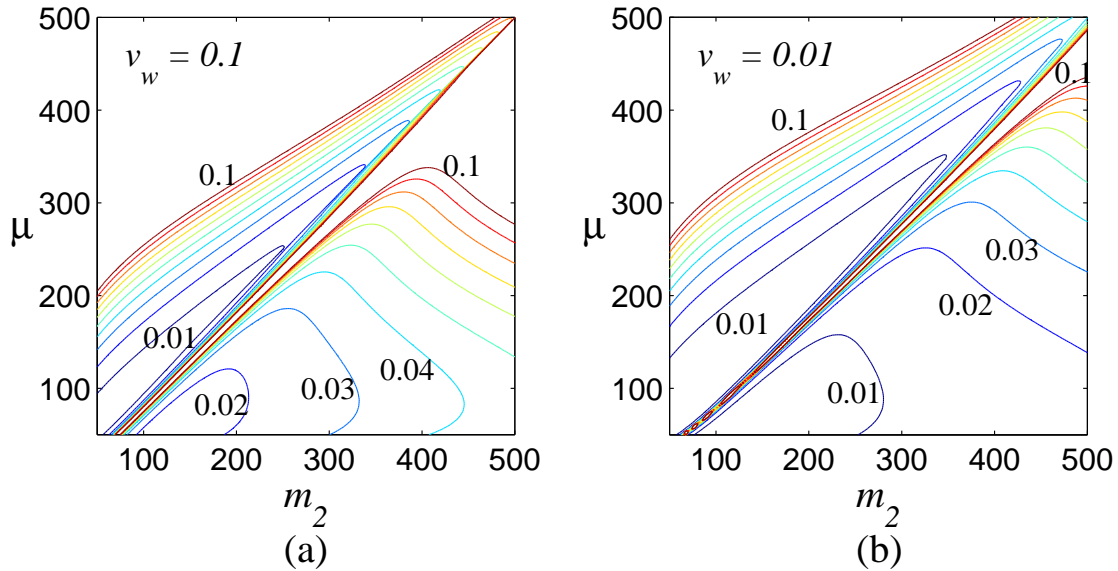}\\[-6mm]
\caption[fig4]{\label{fig4}  Contours of constant CP-violating phase
$\delta_\mu$, corresponding to baryon asymmetry
    $\eta_{\sss B} = 3 \times 10^{-10}$
for (a) $v_w = 0.1$  and (b) $v_w = 0.01$. Mass units are GeV.}
\end{figure}

\section{Conclusions and outlook}

We have presented a detailed analysis of electroweak baryogenesis
in the minimal supersymmetric standard model (MSSM) using
the classical force mechanism (CFM).  We argued that the dominant
baryogenesis source in the MSSM arises from the chargino sector.  We
also commented on a recent controversy regarding the parametric form of
the source appearing in the diffusion equations.  The resolution is
that all different formalisms agree with the parametric form; however
previous authors neglected the particular source considered here for
the linear combination of Higgsinos $H_1+H_2$, on the grounds that it
is suppressed by the top (s)quark Yukawa interactions.  We have shown
that this suppression is quite modest, a factor of order unity,  which
is much milder than the intrinsic suppression suffered by the competing
source $H_1-H_2$, due to the near constancy of the ratio $H_1/H_2$
throughout the bubble wall \cite{MQS,CM}.

Our present work differs in several ways from our earlier published
results using the CFM. First we have presented a treatment in terms of
the physical, kinetic variables characterizing the WKB states rather than
in terms of canonical variables, which are gauge dependent, and in terms
of which the recovery of a gauge independent physical result is not
always transparent. While this is mainly a matter of (considerable!)
convenience, there is also a slight physical difference in our results
due to a slightly different form for equilibrium ansatz corresponding
to each parametrization (see appendix B).
We noted that a definitive determination of the correct form will have
to await the outcome of a more fundamental computation, as will the
correct treatment of `spontaneous' baryogenesis, which rely completely
on the relevant form of the collision integral. Much more importantly
to quantitative changes in our results is that in our treatment in
\cite{CJK} we misidentified the sign of the hypercharge of one of the
Higgsino states.  This prevented us from realizing that the top Yukawa
interactions tend to damp the appropriate combination of Higgsino
currents, $H_1 + H_2$. Here we developed a new set of diffusion
equations where this effect is treated correctly. Thirdly, here we have
considered several different choices for which flavors of squarks are
light compared to the temperature, and found that the one adopted in
\cite{CJK} (all squarks light) is among the less favoured possiblities
for baryogenesis, because of strong sphaleron suppression.

We have given a complete derivation of the CFM formalism, starting
from the basic assumption that the plasma is adequately described by a
collection of  WKB-quasiparticle states in the vicinity of a varying
background Higgs field.  We derived the dispersion relations for
squarks and charginos, and showed how to identify the appropriate
kinetic momentum variable, the physical group velocity, and the force
(see also \cite{KK}).  We pointed out that the force term for the
current combination $H_1 - H_2$, obtained in a recent publication
\cite{HS}, is absent in physical variables and hence this current is
not sourced in the CFM. This is a very sensible result, because the term
in question can always be removed by a canonical transformation, or
equivalently, a field redefinition.  We also derived the diffusion
equations and the source terms appearing in them starting from the
semiclassical Boltzmann  equations for the quasiparticle  states.

We have studied the baryon production efficiency in the MSSM as a function
of the parameters in the chargino mass matrix, the wall width and the wall
velocity.  The dependence on wall velocity is rather complicated, and
intertwined with the dependence on the chargino mass parameters; the
generated asymmetry generically changes sign as a function of $v_w$, but
the value of $v_w$ where this crossing takes place, and the the functional
form of $\eta_B(v_w)$, are quite dependent on mass parameters. However,
for large regions of parameters an asymmetry  $\eta_{10} \equiv 10^{10}
\eta_B$ of the order of several hundred could be created, implying that a
CP-violating angle of $\arg(m_2\mu ) \simeq 10^{-2}$, and in best cases
even $\arg(m_2\mu ) \simeq$ a few$\times 10^{-3}$, suffices for producing the
observed asymmetry of $\eta_{10} \sim 3$ \cite{FKOT}. Such small phases
are consistent with the present limits from the neutron and electron
dipole moment constraints \cite{dmom}.

We finally emphasize that our formalism disagrees in detail with
various other methods of computing the source in the diffusion equations.
This is particularly significant with regard to references \cite{HuSa,HN,
Riotto,CQRVW}, which all claim to be valid in the thick wall regime, where
our method was designed to work. This is troubling, because one
expects that different methods should agree when the same physical
limits are taken. In particular, we have shown that classical force
mechanism does not give rise to a source of the parametric form
$\sim H_1'H_2 - H_1H_2'$, found by references \cite{HuSa,HN,Riotto,
CQRVW,HS} for both squarks and charginos. We do not know a definite
solution to this problem, but a possible
origin for the discrepancy could be that the methods \cite{HuSa,HN,Riotto,
CQRVW} perform an expansion in the mass, or the vacuum expectation value,
divided by temperature (the mass insertion expansion \cite{Riotto}),
before taking the gradient expansion.  In the WKB approach on the other
hand the background is treated in a mean field approximation and one performs
the gradient expansion around this classical background. In other words, in
the WKB-picture the mass insertion expansion has been resummed to infinite
order before the gradient limit is considered. While one can formally expand
the CFM-source resulting from a WKB-analysis in mass over temperature, one
should in general not expect that taking these two limits is commutative.
In particular, quantum reflection is completely absent in the WKB
approximation, but is certainly present in the mass insertion expansion.
The issue of how to properly account for both the semiclassical and
the quantum effects, or to interpolate between them, is certainly worth
further study, and some published results from a work aiming to a
derivation of appropriate semiclassical Boltzmann equations from
first principles can be found in references \cite{JKPb,JKPf}. \\

\section*{Acknowledgement}
We are grateful to Dietrich  B\"odeker, Guy Moore, Tomislav Prokopec and
Kari Rummukainen for many clarifying discussions, constructive comments and
for providing useful insights on various issues related to this work.\\

%
\section*{Appendix A: Collision Terms in Linear Expansion}
%

We show here how the collision integral on the r.h.s. of the Boltzmann
equation gives rise to the terms damping the perturbations from equilibrium.
For illustration, let us consider a two body process with ingoing WKB states
$i$ with four momenta $p_i$ (with $p_{i1}$ corresponding to the distribution
on the l.h.s.\ of the Boltzmann equation) and outgoing WKB states $f$ with
four momenta $p_f$:
\beq
C [f_j] = {1 \over 2 E_{i1}}
      \int_{p_{i2},p_f} |{\cal M}|^2 (2\pi )^4
      {\delta\!\!\!^-}^4 \Bigl (\sum_{l}\hat{p}_l\Bigr )
      {\cal P}[f_j]
\label{twobytwo}
\eeq
where $ |{\cal M}|^2$ is the matrix element calculated between the WKB
states, to first order in derivatives of the background, $\int_p$ means
$\int \rd^3p/2E(2\pi )^3$ and the statistical factor ${\cal P}[f_j]$ is
given by
\beq
{\cal P}[f_i] =  f_{i1} f_{i2} (1\mp f_{f1})(1\mp f_{f2})
                    - f_{f1} f_{f2} (1\mp f_{i1})(1\mp f_{i2}).
\label{stat}
\eeq
We only attempt to compute the collision integrals to the zeroth order
accuracy in gradients, so that the integral measures and $\delta$-functions
are the same as in the usual flat space-time considerations.  Most part of
the derivation consists of manipulating the statistical factor ${\cal
P}[f_i]$.
Inserting the ansatz (\ref{dist}) to (\ref{stat}) and  expanding to the first
order in $\delta\! f$ one gets
\beq
{\cal P}[f_i(\mu_i)+\delta\! f_i]  \simeq {\cal P}[f_i(\mu_i)]
+ f^0_{i1} f^0_{i2} f^0_{f1} f^0_{f2} e^{\beta (E_{i1}+E_{i2})}
\left(
       \frac{\delta\! f_{i1}}{f^0_{i1}} + \frac{\delta\! f_{i2}}{f^0_{i2}}
- \frac{\delta\! f_{f1}}{f^0_{f1}} - \frac{\delta\! f_{f2}}{f^0_{f2}}
\right),
\label{stat1}
\eeq
where $f_i(\mu_i)$'s are the distributions given by the first term in the
ansatz (\ref{dist}) and $f^0_j$'s are the unperturbed distribution functions.
${\cal P}[f_i(\mu_i)]$ is nonzero only for inelastic scatterings, whereas all
reaction channels create nonvanishing collision terms proportional to
$\delta\! f_j$; let us consider these terms first.

The entire collision integral corresponding to $\delta\! f_j$-terms in
(\ref{stat1})can be
written as
\beqa
C[\delta\! f_l] &\simeq & \delta\! f_{i1}(p_1) \Gamma (p_{i1})
      + f^0(p_{i1})\int_{p_{i2}} \delta\! f_{i2}(p_{i2}) a(p_{i1},p_{i2})
      \nn \\ &&\phantom{Hannatytton}
       - \sum_n \int_{p_{fn}} \delta\! f^0_{fn}(p_{fn}) G_{fn}(p_{i1},p_{fn})
      \nn \\
      &\equiv &  \delta\! f_{i1}(p_1) \Gamma (p_{i1}) - \delta\! F(p_{i1})
\label{terms}
\eeqa
The abbreviated notation here is used to highlight the fact that only the
first term is directly proportional to $\delta\! f_{i1}(p_{i1})$, whereas all
the others contain smeared integrals over $\delta\! f_j$-distributions; the
exact forms of the functions $a(p,k)$ and $G_i(p,k)$ are not relevant for
us. $\Gamma (p_{i1})$ on the other hand is the usual thermally averaged
interaction rate, which, neglecting the Pauli-blocking factors, is given
by
\beq
\Gamma (p_{i1}) = \int \frac {\rd^3 p_{i2}}{(2\pi )^3}
                       f^0_{i2}(p_{i2})(v_{\rm rel} \sigma_{i\ra j}).
\label{gammap}
\eeq
where
\beq
(v_{\rm rel}\sigma_{i\ra j}) \equiv \frac{1}{4E_{i1}E_{i2}}\int_{p_f}
                 (2\pi )^4\delta^4(\sum_l p_l) |{\cal M}|^2
\label{vsigma}
\eeq
is the invariant cross section for the process $i\ra f$ multiplied by
the invariant flux (see for example \cite{DK}).

The $\delta\! f\;\Gamma$-term in (\ref{terms}) clearly causes damping away of
kinetic fluctuations, with a momentum dependent relaxation scale given by
the inverse of the rate (\ref{gammap}). The ``noise term'' $\delta\! F$
physically represents the process of further thermal redistribution of the
states which goes on alongside the relaxation of $\delta\! f_i$ to zero.
Thes are random processes which occasionally oppose the relaxation process.
However, while the integrated over $p_{i1}$ moments of $\delta\! F(p_{i1})$
are comparable to the moments of the first term, their naive inclusion to
the moment equations would be incorrect, since kinetic relaxation depends
sensitively on the shape of the entire distribution function. For example,
the condition $\ave{(p_{i1}/E_{i0})(\delta\! f \Gamma - \delta\! F)} = 0$
would lead to vanishing of the (kinetic) relaxation term for the velocity
perturbation in moment equations, whereas in reality the kinetic relaxation
process is halted only if the collision term (\ref{terms}) vanishes
identically for {\it all} momenta.  The effect of the noise terms is further
reduced by the fact that in all elastic channels adding the scatterings
from particles and antiparticles tend to cancel the noise part, while
the contributions to the relaxation terms are equal and add.  In the
diffusion approximation then, the first moments of the part of the
collision term containing $\delta\! f_j$'s are given by
\beqa
\ave{C[\delta\! f_j]} &\simeq& 0 \nn \\
\ave{v_{p_z}C[\delta\! f_j]} &\simeq&
      \ave{v_{p_z}\delta\! f_{i1}(p_z)}\bar \Gamma ,
\label{delfmoments}
\eeqa
where the average $\ave{\cdot}$ is as defined in equation (\ref{average}). In
the case of first moment we have also assumed that $\Gamma(p)$ has only a weak
momentum dependence so that it can be replaced by its thermal average $\bar
\Gamma$.  This is of course the place where we implicitly truncate our momentum
expansion to the first two terms, and it should be a very good approximation.
Adding up the contributions from all possible  channels, one obtains the result
(\ref{Cmoment0}).

Let us next consider the ${\cal P}[f(\mu_j)]$-part of the statistical
factor (\ref{stat1}). Expanding to the first order in $\mu_j$'s one finds
\beq
{\cal P}[f(\mu_j)] \simeq - f^0_{i1} f^0_{i2}
            (\xi_{i1} + \xi_{i2} - \xi_{f1} - \xi_{f2}),
\label{expform}
\eeq
where $\xi\equiv \mu/T$ and we also neglected the Pauli blocking factors in
the final states. This expression obviously vanishes for the elastic channels,
whereas for inelastic channels it gives the contribution
\beq
C[\mu_j] \simeq f^0_{i1}(p_{i1}) \Gamma_i (p_{i1}) \sum_j \xi_j
\label{inel}
\eeq
where $\Gamma_i(p_{i1})$ is an expression analogous to
(\ref{gammap}) and $\sum_j \xi_j$ is the signed sum over the chemical
potentials appearing in (\ref{expform}) such that the term $\xi_{i1}$ has a
positive sign. The first moments of the inelastic collision term in
(\ref{inel}) then are
\beqa
     \ave{C[\mu_j]}  &\simeq& \bar\Gamma_i \sum_j \xi_j
\nn \\
     \ave{v_{p_z} C[\mu_j]}  &\simeq& 0
\label{inelmom}
\eeqa
where
\beq
     \bar \Gamma_i \equiv
         \int_{p_{i1}} f^0_{i1} \Gamma_ i(p_{i1})/N_{i1}
     \label{gammainel}
\eeq
with $N_{i1}\equiv \int \rd^3p f_{i0}'$ (see equation (\ref{average})).
Adding up all inelastic channels affecting a given species $i$, one
arrives to the equation (\ref{Cmoment1}). \\

%
\section*{Appendix B: Equilibrium ansatz in canonical variables}
%

{\bf Gauge invariance.} In previous treatments \cite{JPT,CJK} an ansatz
different from (\ref{disteq}) was adapted for the local equilibrium
function:
\beq
      \tilde{f} (p_c,x) =
{1\over e^{\beta[ \gamma_w(\omega + v_w p_c) - \tilde{\mu}(x)]}
                  \pm 1}  + \delta\! \tilde{f}(p,x),
\label{dist-old-app}
\eeq
where $p_c$ is the canonical momentum. A technical problem with the
canonical variables is that both $p_c$ and $\tilde{\mu}$ are  phase
reparametrization, or ``gauge" variant quantities. This can be seen by
observing that any physical quantity (e.g.\ local number density) is
obtained integrating over the momenta. The integration measure $d^3p_c$
is unchanged by gauge transformation $p_c \rightarrow p_c + \acp$, so
that a system with fixed number density is described by a different
value of $\tilde{\mu}$ in two different gauges.

In the previous treatments \cite{JPT} and \cite{CJK} equations for
physically meaningful quantities were recovered using the condition that
the system be unperturbed from equilibrium far in front of the wall (at
$z \rightarrow \infty$) i.e.
\beq
      \int d^3 p_{\infty} \tilde{f}
    = \int d^3 p_{\infty}
      \frac{1}{e^{\beta [\gamma_w(\omega + v_w p_{\infty})-\mu_{\rm 
phys}]}\pm 1}
      \qquad  \mu_{\rm phys} \rightarrow 0
      \quad {\rm as} \quad z\rightarrow \infty
\label{jpt-bc}
\eeq
where $p_\infty$ is the physical momentum at infinity. For the case of
a fermion with complex mass discussed in section 2, Eqns.\ (\ref{drsimple})
and (\ref{drsimpleR}) give
\beq
     p_{\infty} = p_{c,\infty} + \scp\frac{s\theta'}{2} - \acp
\label{phys-infty}
\eeq
(since $m \rightarrow 0$), where $\acp \equiv \alpha' + \scp \theta'/2$
in the leftchiral sector for example. One then identifies the physical
chemical potential as
\beq
    \mu_{\rm phys} = \tilde{\mu}+v_w\gamma_w(\scp\frac{s\theta'}{2} - \acp).
\label{phys-cp}
\eeq
The equations are then most conveniently rewritten in terms of
$\mu_{\rm phys}$ and $p_{\infty}$, and solved with with the boundary
condition $\mu_{\rm phys}=0$ at $+\infty$. Note that in both \cite{JPT}
and \cite{CJK} a specific gauge was chosen, which can be read off
from the dispersion relations adapted as $\acp=0$ in \cite{JPT}, and
$\acp=\scp\frac{s\theta'}{2}$ in \cite{CJK}. While in the former case
a transformation from the original canonical variables had to be performed
(cf. section 4, page 2962 in \cite{JPT}), the implicit
gauge choice of \cite{CJK} required no such transformation, since
$\mu_{\rm phys}=\tilde{\mu}$ in this gauge. Note in particular that
the terms proportional to the linear combination of scalar fields
$H_1'H_2 - H_2'H_1$ appearing in canonical momenta (\ref{DRsq}),
(\ref{charginoDR}) and (\ref{charginoDR2}), can entirely be absorbed
into the redefinition of $\mu_{\rm phys}$, so they will not provide
new sources even when treating the problem using the canonical
momentum.\\

\noindent{\bf Comparision of the ans\"atze.}  Making use of
(\ref{vgroup}) and the dispersion relation (\ref{engy}) one can
show that the canonical and kinetic momenta are, to linear order in
$\theta'$, related by
\beq
    p_z(1-\scp\frac{s\theta'}{2\tilde \omega })=p_c -\acp ,
\label{kinetic-can}
\eeq
where $\tilde \omega \equiv \sqrt{\omega^2 - p^2_{||}}$. In kinetic
variables the ansatz (\ref{jpt-bc}) may then be written as
\beq
   \tilde{f}(p,x) =
   \frac{1}{e^{\beta [\gamma_w(\omega + v_w p) - \mu_{\rm phys}]} \pm 1}
      + \delta\! \tilde{f}(p,x) +
   \beta v_w \gamma_w \scp \frac{s\theta'}{2}
         \frac{\tilde \omega -p}{\tilde \omega} f' ,
\label{dist-old}
\eeq
where $f'=(1/e^x\pm1)'$ ($x=\beta \omega$).  One can thus identify
the physical chemical potential $\mu_{\rm phys}$ in the canonical
variables with the chemical potential for our physical
WKB-quasiparticles appearing in the ansatz (\ref{dist}). The chemical
potentials in (\ref{dist}) and (\ref{dist-old-app}) are thus only
separated by an unphysical gauge-transform. The distributions differ
however, by a term that cannot be transformed away: to make
(\ref{dist-old-app}) completely agree with (\ref{dist}), one should
have
\beq
    \delta\! \tilde{f}  =  \delta\!f
    - \beta v_w \gamma_w \scp \frac{s\theta'}{2}
                         \frac{\tilde \omega-p}{\tilde \omega} f' .
\label{substitutions}
\eeq
The latter term does not vanish in equilibrium however, so the two
ans\"atze do correspond to two physically different equilibrium conditions.
The difference is only nonzero in the region of the wall however, as
it vanishes when $|m| \ra 0$. Including this term in our Boltzmann
equations would contribute to source term, making it equal to the one
used in \cite{CJK}.

It is clear that the difference between ans\"atze corresponds to which
energy momentum - canonical or kinetic - is the appropriate one to take
as that conserved in the local interactions between particle states
modelled there. We have argued in the main text that the kinetic
momentum has the more direct physical interpretation, and this argument
is backed up by results from a more sophisticated treatment \cite{JKPf,
JKPb}. However, a complete derivation of the transport equations using
the formalism of \cite{JKPf,JKPb} will be needed to settle the issue
unambiguously, while in practice the numerical results are not
particularily sensitive to the difference.\\

\nc{\APJ}[3]   {{\it Ap.\ J.\ }                {{\bf #1} {(#2)} {#3}}}
\nc{\APJL}[3]  {{\it Ap.\ J.\ Lett.\ }         {{\bf #1} {(#2)} {#3}}}
\nc{\APP}[3]   {{\it Astropart.\ Phys.\ }      {{\bf #1} {(#2)} {#3}}}
\nc{\IBID}[3]  {{\it ibid.\ }                  {{\bf #1} {(#2)} {#3}}}
\nc{\IJMP}[3]  {{\it Int.\ J.\ Mod.\ Phys.\ }  {{\bf #1} {(#2)} {#3}}}
\nc{\IJTP}[3]  {{\it Int.\ J.\ Theor.\ Phys.\ }{{\bf #1} {(#2)} {#3}}}
\nc{\JP}[3]    {{\it Jetp.\ Lett.\ }           {{\bf #1} {(#2)} {#3}}}
\nc{\MPL}[3]   {{\it Mod.\ Phys.\ Lett.\ }     {{\bf #1} {(#2)} {#3}}}
\nc{\NAST}[3]  {{\it New Astronomy}            {{\bf #1} {(#2)} {#3}}}
\nc{\NCIM}[3]  {{\it Nuov.\ Cim.\ }            {{\bf #1} {(#2)} {#3}}}
\nc{\NP}[3]    {{\it Nucl.\ Phys.\ }           {{\bf #1} {(#2)} {#3}}}
\nc{\PR}[3]    {{\it Phys.\ Rev.\ }            {{\bf #1} {(#2)} {#3}}}
\nc{\PRL}[3]   {{\it Phys\ Rev.\ Lett.\ }      {{\bf #1} {(#2)} {#3}}}
\nc{\PL}[3]    {{\it Phys.\ Lett.\ }           {{\bf #1} {(#2)} {#3}}}
\nc{\PREP}[3]  {{\it Phys\. Rep.\ }            {{\bf #1} {(#2)} {#3}}}
\nc{\PHYS}[3]  {{\it Physica }                 {{\bf #1} {(#2)} {#3}}}
\nc{\PTP}[3]   {{\it Prog.\ Theor.\ Phys.\ }   {{\bf #1} {(#2)} {#3}}}
\nc{\RMP}[3]   {{\it Rev.\ Mod.\ Phys.\ }      {{\bf #1} {(#2)} {#3}}}
\nc{\RPP}[3]   {{\it Rep.\ Prog.\ Phys.\ }     {{\bf #1} {(#2)} {#3}}}
\nc{\SJNP}[3]  {{\it Sov.\ J.\ Nucl.\ Phys.\ } {{\bf #1} {(#2)} {#3}}}
\nc{\SPJETP}[3]{{\it Sov.\ Phys.\ JETP\ }      {{\bf #1} {(#2)} {#3}}}
\nc{\YF}[3]    {{\it Yad.\ Fiz.\ }             {{\bf #1} {(#2)} {#3}}}
\nc{\ZETP}[3]  {{\it Zh.\ Eksp.\ Teor.\ Fiz.\ }{{\bf #1} {(#2)} {#3}}}
\nc{\XP}[3]    {{\it Z.\ Phys.\ }              {{\bf #1} {(#2)} {#3}}}


\vfill\eject
\baselineskip=20pt

\end{document}